\input{aipcheck}

\newcommand\ffcaption[1]{%
\footnotesize{\bf \footnotesize \refstepcounter{figure}%
\figurename~\thefigure.}{\ #1}}
\documentclass[
    final,            
  ]
  {aipproc}

\usepackage{floatflt}

\layoutstyle{6x9}
\usepackage{epsf}
\usepackage{graphicx}

\usepackage{amsmath}
\usepackage{amssymb}
\usepackage{epsfig} 
\usepackage{rotating}

\def\to{\rightarrow}

\def\te{\tilde e}

\def\tu{\tilde u}

\def\ttau{\tilde \tau}

\def\tell{\tilde\ell}

\def\tw{\widetilde W}
\def\tz{\widetilde Z}
\def\alt{\stackrel{<}{\sim}}

\def\tc{\tilde c}

\newcommand{\bi}{\begin{itemize}}
\newcommand{\ei}{\end{itemize}}
\newcommand{\be}{\begin{equation}}
\newcommand{\ee}{\end{equation}}
\newcommand{\nosusy}{{{SUSY}\hspace*{-0.5cm}\Big/\hspace*{0.5cm}}}

\begin{document}

\title{SUSY: THEORY STATUS IN THE LIGHT OF  EXPERIMENTAL CONSTRAINTS}

\author{Alexander Belyaev}{
  address={Department of Physics, Florida State University\\ 
Tallahassee, FL 32306, USA\\
E-mail: {belyaev@hep.fsu.edu}}
}

%

\begin{abstract}
Supersymmetry remains compelling theory over 30 years in spite of lack of its
discovery. It could be already   near the corner our days, therefore  present
and upcoming experiments  are crucial  for  constraining or even discovery
of the supersymmetry.
\end{abstract}

\maketitle

\section{Introduction}
Since  1970's the Standard Model~(SM)  based on local
gauge invariance  principle explains all experimental data with a good
precision. It  is based on $SU(3)_c\times SU(2)_L\times U(1)_Y$ 
non-Abelian Yang-Mills type gauge theory spontaneously broken to 
{$SU(3)_c\times U(1)_Y$} group. 

Interestingly, at about the same time,  when SM has been established,
the first ideas of Supersymmetry appeared independently 
around the world: 
``Extension of the algebra of Poincar\'e group generators  and violation of P
invariance" of Golfand and Likhtman~(1971)~\cite{Golfand:1971iw}, 
``Dual theory for free fermions" of Ramon~(1971)~\cite{Ramond:1971gb},
``Quark model of dual pions" of Neveu and Schwarz~(1971)~\cite{Neveu:1971iv}, 
``Is the neutrino a Goldstone particle?" of Volkov and Akulov~(1973)~\cite{Volkov:1973ix}. 
One should mention separately the
work of  Wess and Zumino, ``Supergauge transformations in 
four-dimensions"~(1974)~\cite{Wess:1974tw},
where the first 4D supersymmetric quantum field theory has been
formulated,   which has caused the great escalation of the interest of physics community 
in  supersymmetric theories.

Supersymmetry is the symmetry, which relates bosons  and fermions and
transforms them one into another by acting with supersymmetric fermionic
generators $Q$:
\begin{equation}
 Q|\mbox{BOSON}\rangle =|\mbox{FERMION}\rangle, \ \ \ 
 Q|\mbox{FERMION}\rangle =|\mbox{BOSON}\rangle
\label{eq-q}
\end{equation}
\begin{floatingfigure}{5cm}
\includegraphics[width=5cm]{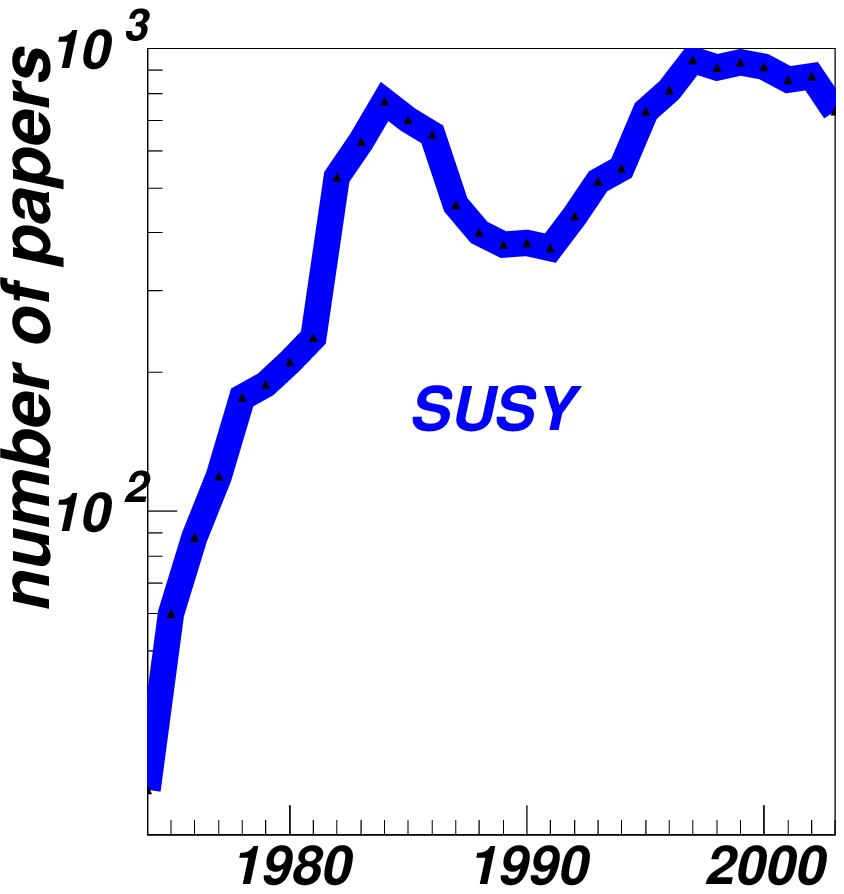}
\ffcaption{
The number of papers published on Supersymmetry versus time}
\label{fig-susy-papers}
\end{floatingfigure}
 One can consider Supersymmetry as one of the most promising attempts to
understand and explain the origin of the fundamental difference between the two classes
of particles -- bosons and fermions. This is already itself quite a  reason for the
theoretical  attractiveness of the Supersymmetry. 
Supersymmetry predicts
"mirror" particles to those of the SM, which we know. 
These supersymmetric
partners should differ from their SM partners by spin $1/2$ and have the same mass if the
Supersymmetry is unbroken. 
Since it is not the case, and we do not observe this
"mirror" supersymmetric  world, 
one can conclude that Supersymmetry either does not
take place  or it is broken. This review is  about the second option, 
since -- as author will try to convince you --
there is a big chance that Supersymmetry did realized
in nature.

Supersymmetry was invented more then 30 years ago but still had not been
found yet. However, people are publishing thousands of papers 
on Supersymmetry per year till present time
and the annual number of the papers does not go down
as one can see from Fig.~1. What makes SUSY being so
attractive  in spite of the 30 years unsuccessful hunting for it?! There are
several fundamental reasons for this.

One of them is that  super-Poincar\'e group -- the group of the Supersymmetry --
contains the most fundamental set
of space-time symmetries as was shown by Haag, Lopusanski and  Sohnius in
1975~\cite{Haag:1974qh}. It contains space-time symmetries of the Poincar\'e
group and includes in addition the  supersymmetry transformation, linking
therefore bosons and fermions. It would be strange if the nature  did not use
this complete set of symmetries.

Besides this, the requirement for supersymmetric transformation to be {\it
local} yields  spin-2 massless gauge filed, the graviton,
mediating gravitational 
interactions~\cite{Nath:1975nj,Freedman:1976xh,Deser:1976eh}. Therefore,
Supersymmetry could provide unification of all forces in Nature, including
gravity.

Supersymmetry also allows to include fermions into Supestring theories which
solve the problem of non-renormalizability of the gravitational theory and pretend
to solve the ultimate goal of construction of theory of everything.

Eventually, the ideas of Supersymmetry --  yet a pure theoretical invention --
survived  for  more than 30 years because our common belief in unification.

\begin{figure}
\epsfxsize=6.cm\epsffile{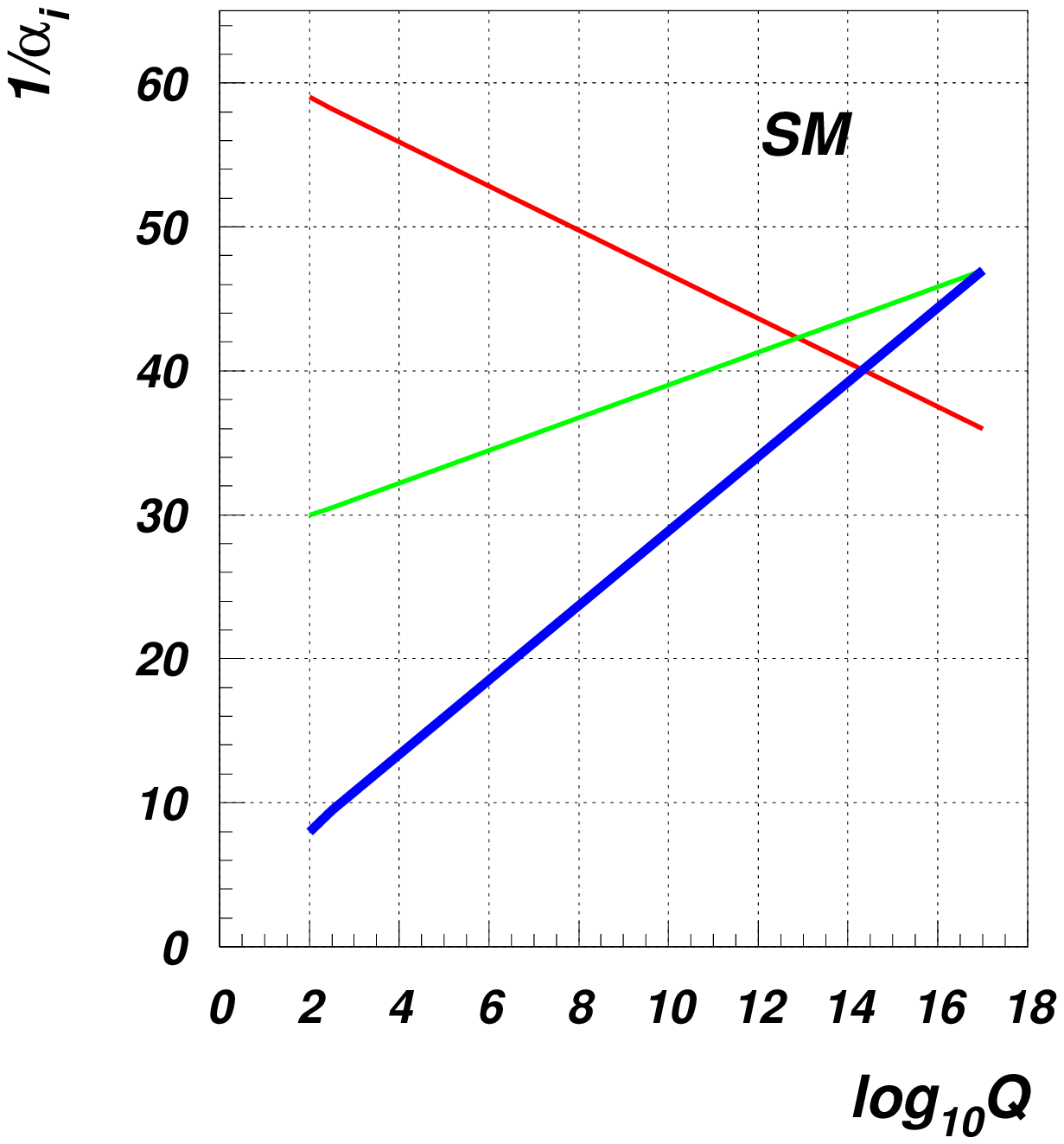}
\epsfxsize=6.cm\epsffile{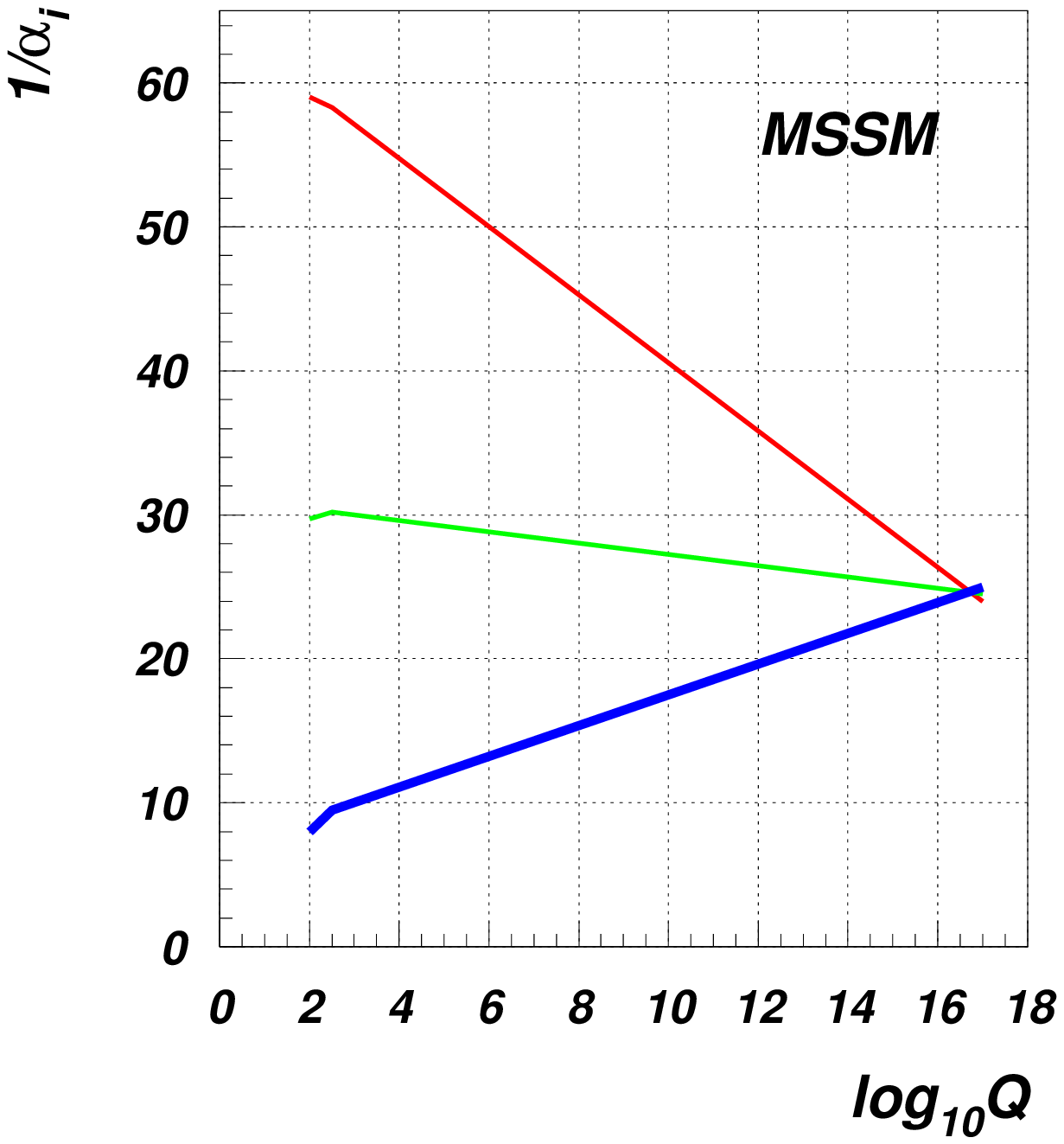}
\caption{\label{fig-susy-couplings}
Evolution of the inverse coupling in the SM (left) and MSSM (right).
}
\end{figure}

Miraculous consequence of the Supersymmetry is the prediction  of the  {\it
unification of the gauge couplings} and solution of the {\it gauge hierarchy
problem } which are central theoretical problems of the SM.  
In SM one can
calculate the evolution of the gauge coupling with the energy scale by running 
respective $\beta$-functions. Evolution of $SU(3)$, $SU(2)$ and $U(1)$ 
coupling from electroweak~(EW) scale up to the $~10^{15}-10^{16}$GeV (GUT) scale
does not give point-like unification of electromagnetic, electroweak and 
strong forces in case of the SM (Fig.~\ref{fig-susy-couplings}, left frame). In case of
minimal Supersymmetric extension of the SM (MSSM), SUSY  particles modify 
beta-functions at about 1 TeV scale and all three couplings do meet together
around 
$M_{GUT}~\sim 10^{16}$~GeV~(Fig.~\ref{fig-susy-couplings}, right).

It is worth to stress that   attractive features of the supersymmetry 
mentioned above, which solve
principal theoretical problems  of the SM are only {\it consequences}
(which has been derived many years after the SUSY has been formulated!)
of the supersymmetry. 
This fact makes SUSY really attractive from the theoretical point of view.

\section{Supersymmetric generalization of the Standard Model}

This overview is aimed at giving the  current  status
of  SUSY in the light of present phenomenological and experimental constraints. 
Therefore the detailed review of basics of  Supersymmetry 
(see
{\it e.g.} 
~\cite{Muller-Kirsten:1986cw,Wess:1992cp,Drees:1996ca,Lykken:1996xt,Martin:1997ns,%
Polonsky:2001pn})
is outside of the scope of this paper
and will be very brief here.

Supersymmetry relates bosons and fermions, which transform
one into another under the action of fermionic generators of supersymmetry algebra
$Q$ and $\overline{Q}$. Therefore the Poincar\'e algebra  has to include commutators of
4-momentum and angular momentum operators $P_\mu$ and $M_{\mu\nu}$
with the new fermionic generators $Q_\alpha$ and $\overline{Q}_{\dot\alpha}$
as well as the anticommutator of $Q_\alpha$ and $\overline{Q}_{\dot\alpha}$:
{\small
\begin{equation}
\big[ P^\mu,Q_\alpha \big]= 0, \ \ \
\big[ M^{\mu\nu},Q_\alpha \big] = -i(\sigma^{\mu\nu})^\beta_\alpha Q_\beta, \ \ \ 
\big\{ Q_\alpha,Q_\beta \big\}=  0,  \ \ \
\big\{ \overline{Q}_\alpha,\overline{Q}_{\dot\beta} \big\} = 0, \ \ \
\big\{ Q_\alpha,\overline{Q}_{\dot\beta} \big\} = 2\sigma^\mu_{\alpha{\dot\beta}} P_\mu \ \    
\label{eq:susy-comm}
\end{equation}
}
One can see from (\ref{eq:susy-comm}) that consequent application of two supersymmetry
transformations leads to the usual space-time translation generated by $P^\mu$,
which tells us that Supersymmetry is a space-time symmetry. 
Furthermore, making SUSY local, one obtains General 
Relativity, or theory of gravity, or supergravity~\cite{Nath:1975nj, Freedman:1976xh, Deser:1976eh}.

In the simplest case, there is  one pair of $Q$ and $\overline{Q}$ ($N=1$
supersymmetry in contrast with extended supersymmetries with $N \ge 2$). 
$N=1$ is the case of our focus here
in constructing a supersymmetric version of the Standard Model.

{\it Superfield formalism} is the most convenient way for constructing supersymmetric lagrangians.
Ordinary fields in this formalism are being extended to {\it superfields}
which are functions of space-time coordinates and
anti-commuting Grassmann coordinates $\theta^\alpha$ and  $\bar\theta^{\dot\beta}$
($\alpha, \dot\beta = 1,2$). 
There are two kinds of superfields participating in construction of SUSY
lagrangian: {\it chiral superfield}, $\Phi$ 
and 
{\it vector  superfield}, $V$.  Chiral superfield contains chiral supermultiplet: 
$\Phi \sim (A,\psi,F)$,
where $A(x)$ is the complex scalar field, $\psi(x)$ is the complex Weyl spinor and $F(x)$ 
is the auxiliary scalar field, which is being eliminated by equations of 
motions and which is needed to close the supersymmetric algebra.
Vector superfield  contains
 $V \sim (\lambda,v_\mu,D)$, where $\lambda$ is a Weyl fermion, 
$v_\mu$
is the vector (gauge) field and $D$ is the auxiliary field.
When  constructing SUSY Lagrangian it is convenient to define 
 {\it field strength tensor}
(an analog of $F_{\mu\nu}$ in the non-supersymmetric field theory):
\\
$W_\alpha = -\frac14 \overline{D}^2 e^V D_\alpha e^{-V}, \ \ \mbox{and} \ \ \
\overline{W}_{\dot\alpha} = -\frac14 D^2 e^V \overline{D}_{\dot\alpha} e^{-V}$
\\
where $D$ and $\bar{D}$ are covariant derivatives:
$D^\alpha=
           -\partial^\alpha		+i(\bar{\theta}\bar{\sigma}	)^{\mu\alpha}      
	   \partial_\mu, \ \ 
 \bar{D}^{\dot\alpha}=
       \bar{\partial}^{\dot\alpha}	-i(\bar{\sigma}\theta		)^{\mu\dot\alpha}
\partial_\mu$, 
\\
and $\sigma^\mu=(1,\sigma^i), \ \ \bar\sigma^\mu=(1,-\sigma^i)$,
are Pauli matrices ($i=1-3$).

The MSSM Lagrangian should consists of two parts  ---
SUSY generalization of the Standard Model and SUSY-breaking part
(since we do not observe exact  SUSY):
${\mathcal L}_{MSSM}={\mathcal L}_{SUSY}+{\mathcal L}_{breaking}$,
with  ${\mathcal L}_{SUSY}={\mathcal L}_{Gauge}+{\mathcal L}_{Gauge-matter}+{\mathcal L}_{matter}=$
\begin{eqnarray}
\sum_{Gauge}\frac{1}{4}
\left(\int d^2\theta \ Tr W^\alpha W_\alpha + h.c.\right) 
+
\sum_{Matter}^{}\int d^2\theta d^2 \bar\theta \
\Phi^\dagger_ie^{\displaystyle \sum_{Gauge} g_j\hat V_j }\Phi_i 
+
\int d^2\theta{\cal W}
\end{eqnarray}
where ${\cal W}$ is a  superpotential,  which should be invariant
under the group of symmetry of a particular model and has the following general form:
\be
\int
d^2\theta
 ~[\lambda_i \Phi_i + \frac{1}{2}m_{ij}\Phi_i \Phi_j + \frac{1}{3} y_{ijk}
\Phi_i \Phi_j \Phi_k] + h.c.
\ee

The general form of written above superpotential does not forbid
violation of  baryon ($B$)  and lepton ($L$) numbers. 
The simultaneous presence of both
$B$- and $L$- violating interactions  should be forbidden since it 
leads to the  fast proton decay.  To avoid  this problem 
one can impose a new discrete symmetry, called $R$-parity: $R = (-1)^{3(B-L)+2S}$,
where $S$ is the spin of the particle. All ordinary Standard Model particles have $R=1$,
the superpartners have $R=-1$. 
$R$-parity implies that the superparticles can be produced only in pairs
and that  the lightest supersymmetric
particle (LSP) is {\it stable}, which is a very welcome crucial point making
LSP the best cold dark matter candidate.
In the MSSM one assumes that $R$-parity is conserved.

When constructing MSSM -- 
minimal supersymmetric extension of the Standard 
Model -- one should note that numbers of degrees of freedom of bosons and fermions should be the same,
{\it and} bosonic and fermionic superpartners should have the same quantum numbers.
Looking at the quantum numbers of the SM particles one concludes that
{\it new fermions} -- gauginos (gluino, wino, zino, photino and two higgsinos)
-- should be added to be the superpartners of {\it known bosons}
and 
{\it new bosons (scalars)} -- squarks and sleptons -- 
should be added to be the superpartners of {\it known fermions}.

The important feature of the supersymmetric models is the enlarged Higgs sector.
In particular, Higgs sector of the  MSSM has the additional Higgs doublet.
One of the reasons for it is  that
in supersymmetric model higgsino has the contribution to the gauge
anomaly, therefore one needs two higgsinos with opposite hypercharges
to realize the cancellation of the gauge anomaly. Hence one needs
{\it two} Higgs doublets with opposite hypercharges to make such a cancellation possible.
Another reason for the introduction of additional Higgs doublet is to give masses for both 
up- and down- type quarks. Conjugated Higgs field cannot  be used 
in MSSM (like it was done in SM) since  $\Phi\Phi\Phi^\dagger$ 
form (which would appear in this analogy)
is not  chiral  superfield and therefore is not supersymmetric.

Table~\ref{tab:mssmprt} summarizes the particle contents and the quantum numbers of the MSSM.

{\small
\begin{ltxtable}
\hspace*{0.5cm}
\begin{rotate}{90}
\begin{minipage}{10cm}
{\bf
\hspace*{-2.5cm}Higgs\hspace*{0.5cm} Matter\hspace*{1.0cm}
Gauge
}
\end{minipage}
\end{rotate}
\begin{tabular}{lllc}
Superfield & \ \ \ \ \ \ \ Bosons & \ \ \ \ \ \ \ Fermions &
$SU_C(3),SU_L(2),U_Y(1)$ \\ \hline \hline 
${\bf G^a}$   & gluon \ \ \ \ \ \ \ \ \ \ \ \ \ \ \  $g^a$ &
{ gluino $ \ \ \ \ \ \ \ \ \ \ \ \ \tilde{g}^a$} & (8,1,0) \\ 
${\bf V^k}$ & Weak \ \ \ $W^k$ \ $(W^\pm, Z)$
 & 
{ wino, zino \ $\tilde{w}^k$ \ $(\tilde{w}^\pm, \tilde{z})$} & (1,3,0) \\ 
${\bf V'}$   & Hypercharge  \ \ \ $B\ (\gamma)$ & 
{bino \ \ \ \ \ \ \ \ \ \ \ $\tilde{b}(\tilde{\gamma })$}  & (1,1,0) \\ 
\hline 
$\hspace*{-0.15cm}\begin{array}{l}{\bf L_i} \\  {\bf E_i} \end{array}$ &
{sleptons
\ $\left\{\begin{array}{l} 
\tilde{L}_i=(\tilde{\nu},\tilde e)_L \\ \tilde{E}_i =\tilde e_R \end{array} \right. $}
 & leptons \ $\left\{ \begin{array}{l}
L_i=(\nu,e)_L
\\ 
E_i=e_R \end{array} \right.$ & 
$\begin{array}{c}  (1,2,-1)\\(1,1,2) \end{array} $ \\ 
$\hspace*{-0.15cm}\begin{array}{l} {\bf Q_i} \\ {\bf U_i} \\ {\bf D_i}
\end{array}$ & 
{  squarks \ $\left\{ \begin{array}{l}
\tilde{Q}_i=(\tilde{u},\tilde d)_L \\ \tilde{U}_i =\tilde u_R 
\\
\tilde{D}_i =\tilde d_R\end{array}  \right. $}
 & quarks \ $\left\{
\begin{array}{l} Q_i=(u,d)_L \\ U_i=u_R^c \\ D_i=d_R^c \end{array}
\right.$ & 
$\begin{array}{c} (3,2,1/3)\\ (\bar{3},1,-4/3)\\(\bar{3},1,2/3) \end{array}$ \\ \hline
$\hspace*{-0.15cm}\begin{array}{l} {\bf H_1} \\ {\bf H_2}\end{array}$ &
Higgses \ $\left\{
\begin{array}{l} H_1 \\ H_2 \end{array}  \right. (h,H,A,H^\pm)$ & 
{higgsinos \ $\left\{
 \begin{array}{l} \tilde{H}_1 \\ \tilde{H}_2 \end{array} \right.(\tilde{h_1},\tilde{h_2},\tilde{h}^\pm)  $} &
$\begin{array}{l} (1,2,-1) \\ (1,2,1) \end{array} $ 
\end{tabular}
\renewcommand{\baselinestretch}{1}
\caption{\label{tab:mssmprt}
MSSM particle contents and respective $SU(3)\times SU(2)\times U(1)$ quantum numbers}
\end{ltxtable}
}

The SUSY breaking sector is one  of the most essential in MSSM model.
We do not observe degenerate states of particles and anti-particles,
which unavoidably leads to conclusion  that  SUSY should be broken.
 The implementation
of SUSY breaking
should not spoil the cancellation of the quadratic divergences i.e. 
SUSY should be broken {\it softly}. The most general form of the soft SUSY part should include 
operators with mass dimension $\le 4$~\cite{Girardello:1981wz}:
\be
- {\mathcal L}_{soft  } = \sum_{scalars} m_{ij} A^*_i A_j +
\sum_{gauginos} M_a \left( \lambda_a \lambda_a + \bar\lambda_a \bar\lambda_a \right)
 + \sum_{i,j,k} A_{ijk} \lambda_{ijk} A_i A_j A_k + \mu B H_1 H_2
\ee
This general form ${\mathcal L}_{soft}$, which includes many parameters, 
should be properly constrained in order to address the experimental  non-observation
of flavor and CP-violating processes.

It is known that  none of the fields of the MSSM can develop nonzero 
vacuum expectation value (v.e.v.)
to  break SUSY without spoiling gauge invariance. 
In the  most common scenarios SUSY breaking occur in the {\it hidden sector}
and propagates to the {\it visible sector} via {\it messengers}. 
There are several known mechanisms to mediate SUSY breaking from hidden to visible sector:
gravity mediation (SUGRA), gauge mediation (GMSB), anomaly mediation (AMSB) and 
gaugino  mediation (inoMSB).

In SUGRA scenario~\cite{Chamseddine:1982jx,Barbieri:1982eh,Hall:1983iz} the
hidden sector communicates with visible one via gravity, leading to the SUSY breaking scale
$M_\nosusy \sim m_{3/2}$, where  $m_{3/2}$ is the gravitino mass. Scalar masses $m_{ij}$,
gaugino masses $M_a$ and trilinear couplings  are proportional to $m_{3/2}$ but can be
non-universal in general. In this case one should properly address flavor and CP problem.
In minimal SUGRA scenario the universality hypothesis of equal boundary conditions at the
GUT scale greatly reduces the number of parameters down to four: $m_0$ -- the common scalar
mass,  $m_{1/2}$ -- the common gaugino mass, $A_0$ and $B$ -- the common tri- and bilinear
soft supersymmetry breaking parameters at the GUT scale and   allows to  solve
the  flavor and CP problem.

In the case of gauge mediation, messenger is not gravity but superfield $\Phi$ that couples to
hidden sector and SM fields via gauge interactions~\cite{Dine:1993yw,Dine:1994vc,Dine:1995ag}. 
In this
scenario gaugino masses are generated at 1-loop level: $M_a={g_a^2\over{16 \pi^2}}{\langle
F_S\rangle\over M}$ ($M$ is the messenger mass scale, while $\langle F_S\rangle$ is the v.e.v. of
singlet scalar superfield of the hidden sector), while sfermion masses are generated at 2-loop
level: $\tilde{m}^2_A=2\sum C^A_a\left({g_a^2\over{16 \pi^2}}\right)^2\left({\langle
F_S\rangle\over M}\right)^2$. Remarkable feature of this scenario is that Gravitino is LSP
since $m_{\tilde G}\sim {{\langle F_S\rangle}\over M}\cdot{M\over M_{PL}}$ is suppressed by Plank Mass.
Another specific feature of this scenario is that soft masses are correlated 
to the respective gauge couplings.

\begin{floatingfigure}{8cm}
\hspace*{-0.5cm}\includegraphics[width=8cm]{%
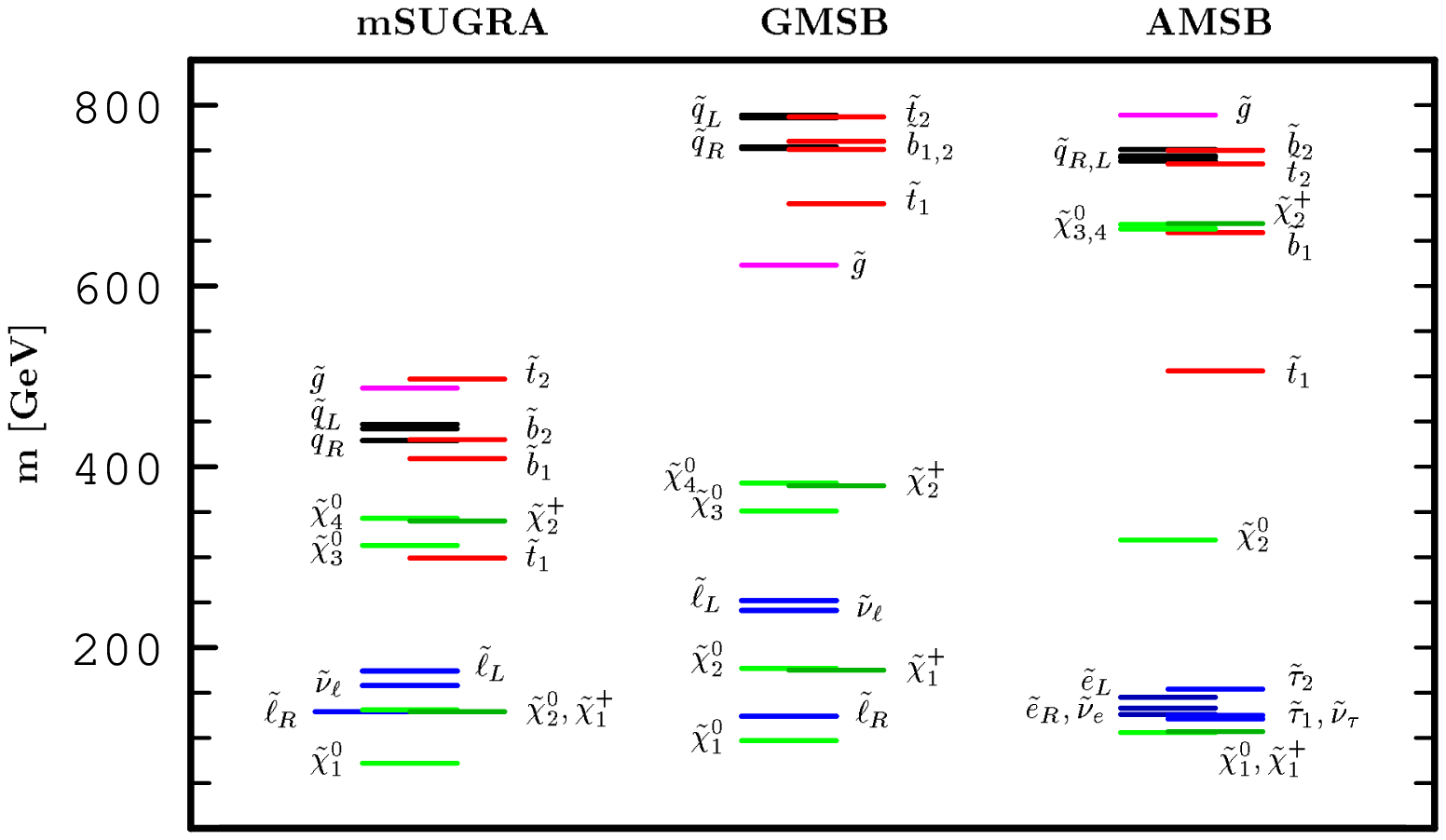}
\ffcaption{Sample spectra for SUSY particles for SUGRA, GMSB and AMSB 
scenarios~\cite{Aguilar-Saavedra:2001rg}}
\label{fig:susy-spectrum}
\end{floatingfigure}

Anomaly mediation as well as Gaugino mediation scenarios are based on the 
paradigm of extra-dimensional brane world. 
Anomaly mediated SUSY breaking~\cite{Randall:1998uk,Giudice:1998xp}  is generated due to
conformal (super-Weyl) anomaly at 1-loop level
while SUGRA mediation effects are exponentially suppressed 
since hidden and visible sectors reside on different branes.
In gaugino mediated scenario~\cite{Kaplan:1999ac,Chacko:1999mi}, again, 
hidden and visible sectors are  on different branes
while gravity and gauge fields are propagate in the bulk and directly 
couple fields on both branes.
As a result, gauginos acquire mass but scalar masses are suppressed 
and may be neglected at the GUT scale.

The mechanism of SUSY breaking determines the sparticle spectrum and therefore is crucial for the SUSY
phenomenology and particle searches. Fig.~3 
from~\cite{Aguilar-Saavedra:2001rg}
presents the sample spectra for three of four models mentioned above.

\section{Hunting for Supersymmetry}

{\bf General strategy}\\
The search for  the  weak scale Supersymmetry 
is one of the prime objectives
of present and future  experiments.

As we discussed above, supersymmetric models can be classified
by the mechanism for communicating SUSY breaking from the hidden
sector to the observable sector.
Among those scenarios
SUGRA  could be considered  as
the most conservative, since it  requires
neither extra dimensions nor new messenger fields while gravity
does exist. 

The so-called {\it minimal} supergravity (mSUGRA) model  has traditionally been
the most popular choice for phenomenological SUSY analyses. In mSUGRA, it is
assumed that the MSSM is valid  from the weak scale all the way up to the GUT
scale  $M_{GUT}\simeq  10^{16}$ GeV, where the gauge couplings unify.  For this
model a simple choice of K\"ahler metric and gauge kinetic function led to {\it
universal} soft SUSY breaking scalar masses ($m_0$),  gaugino masses
($m_{1/2}$) and $A$-terms ($A_0$) at $M_{GUT}$. This assumption of universality
leads to the phenomenologically motivated suppression of  flavor changing
neutral current (FCNC) processes.  However, there is no known physical
principle which gives rise to the desired form of  K\"ahler metric and gauge
kinetic function, and in general non-universal soft breaking mass parameters at
the GUT scale  are expected~\cite{Kaplunovsky:1993rd,Soni:1983rm}.  In
addition,  quantum corrections would  lead to large deviations from
universality at the tree-level~\cite{Choi:1997de}.

 We will also require that electroweak symmetry is broken radiatively (REWSB),
allowing us to fix the magnitude (not the sign) of the superpotential Higgs
mass term $\mu$  to obtain the correct value of $M_Z$.  The bilinear soft
supersymmetry breaking (SSB) parameter $B$ is usually traded for $\tan\beta$
(the ratio of Higgs field vacuum expectation values).  Thus, the parameter set
\begin{equation}
m_0,\ m_{1/2},\ A_0,\ \tan\beta ,\ \ {\rm and}\ \ sign(\mu )
\end{equation}
completely determines the spectrum of supersymmetric matter and Higgs fields.
The results  presented in this review  are  based on
ISASUGRA program, the part of the ISAJET~\cite{Paige:2003mg}
package, which calculates the SUSY particle mass spectrum.
ISASUGRA includes 
full one-loop radiative corrections
to all sparticle masses and Yukawa couplings, 
and minimizes the scalar potential using the
renormalization group improved 1-loop effective potential 
including all tadpole contributions, evaluated at an 
optimized scale choice which accounts for leading two loop terms.
Working in the $\overline{DR}$ regularization scheme, the weak scale values of
gauge and third generation Yukawa couplings are evolved via 2-loop RGEs to
$M_{GUT}$. At $M_{GUT}$, universal SSB boundary conditions are imposed, and all
SSB masses along with gauge  and Yukawa couplings are evolved to the weak scale
$M_{weak}$.  Using an optimized scale choice
$Q_{SUSY}=\sqrt{m_{\tilde{t}_L}m_{\tilde{t}_R}}$,  the RG-improved one-loop
effective potential is minimized and the entire spectrum of SUSY and Higgs
particles is calculated.
There is generally a good agreement 
between Isajet spectrum  and  publicly available
SoftSUSY~\cite{Allanach:2001kg},
Spheno~\cite{Porod:2003um} and 
Suspect~\cite{Djouadi:2002ze} codes,
as detailed in Ref.~\cite{Allanach:2003jw}.

Once the SUSY and Higgs masses and mixings are known, then one should
calculate observables and compare them against experimental constraints.
The most important of these are:
cold dark matter (CDM) bounds as well as direct and indirect CDM
search experiments;
bounds from various collider experiments (including the future ones)
-- LEP, TEVATRON, LHC, NLC;
rare decay processes such as  $b\to s\gamma$, $B_S\to\mu^+\mu^-$, 
$\mu\to e\gamma$;
electric and dipole moments e.g. 
$\delta a_\mu$ -- anomalous magnetic muon moment;
measurement of the proton life time related to SUSY GUTs
theories.

{\bf Cold dark matter bounds.}\\
Among the most important experiments for SUSY search are those,  which confirmed
the evidence of  cold dark matter (CDM) in the Universe and aimed for
direct/indirect search of CDM candidate.
The most direct evidence for CDM in the Universe comes from 
observations of galactic rotation curves.   Also  binding of galaxies in
clusters, matching observations of large scale structure with simulations, 
gravitational microlensing,  baryonic density of the Universe as  determined by
Big Bang nucleosynthesis,  observations of supernovae in  distant galaxies, and
measurements of anisotropies in the cosmic microwave background radiation (CMB)
can be considered as a confident confirmation of CDM (for reviews
see e.g.~\cite{cdm-freedman,cdm-nanopoulos}).
In particular, the  Wilkinson Microwave Anisotropy
Probe~(WMAP)~\cite{Spergel:2003cb,Bennett:2003bz}  collaboration  has extracted
a variety of cosmological parameters from fits to precision measurements of the
CMB radiation. It was found that
\begin{equation}
\Omega_{total}=1.02\pm 0.02 
\end{equation}
where $\Omega_{total}=\rho_{total}/\rho_c$,  $\rho_{total}$ is the
matter-energy density of the Universe, $\rho_{c}$ ($G_N$ is the Newton constant)
the critical density  of the closure of the Universe, defined within
Friedman-Robert-Walker (FRW) framework. $H$ is the Hubble parameter usually
defined through the  rescaled Hubble parameter $h$ as $H=100h$~Km/(s Mpc). So,
with a good precision, Universe has been measured to be flat.  The properties
of the Universe are characterized by  the density of baryons ($\Omega_b$),
matter density  ($\Omega_m$), vacuum energy  ($\Omega_\Lambda$) and the
expansion rate ($h$) which are measured to be:
\begin{equation}
\Omega_b=0.044\pm 0.004, \ \ \ \
\Omega_m=0.27\pm 0.04,   \ \ \ \
\Omega_\Lambda=0.73\pm 0.04, \ \ \ \
h=0.71^{+0.04}_{-0.03} .
\end{equation}

From the WMAP results, one derives the following value for CDM:
\begin{equation}
\label{eq:wmap}
\Omega_{CDM}h^2=\Omega_{m}h^2-\Omega_{b}h^2 = 0.1126^{+0.0081}_{-0.0090}
( {^{+0.0161}_{-0.0181}} )  \mbox{ at 68(95)\% CL}.
\end{equation}
The energy content of the Universe according to WMAP data
is schematically  presented in Fig.~4.
There exists a number of hypothetical 
candidate elementary particles to fill the role of CDM.
A particularly attractive candidate for CDM is the lightest neutralino in
$R$-parity conserving supersymmetric models~\cite{Goldberg:1983nd,Ellis:1983ew}.

\begin{floatingfigure}[l]{6.5cm}
\epsfxsize=6.2cm\epsffile{%
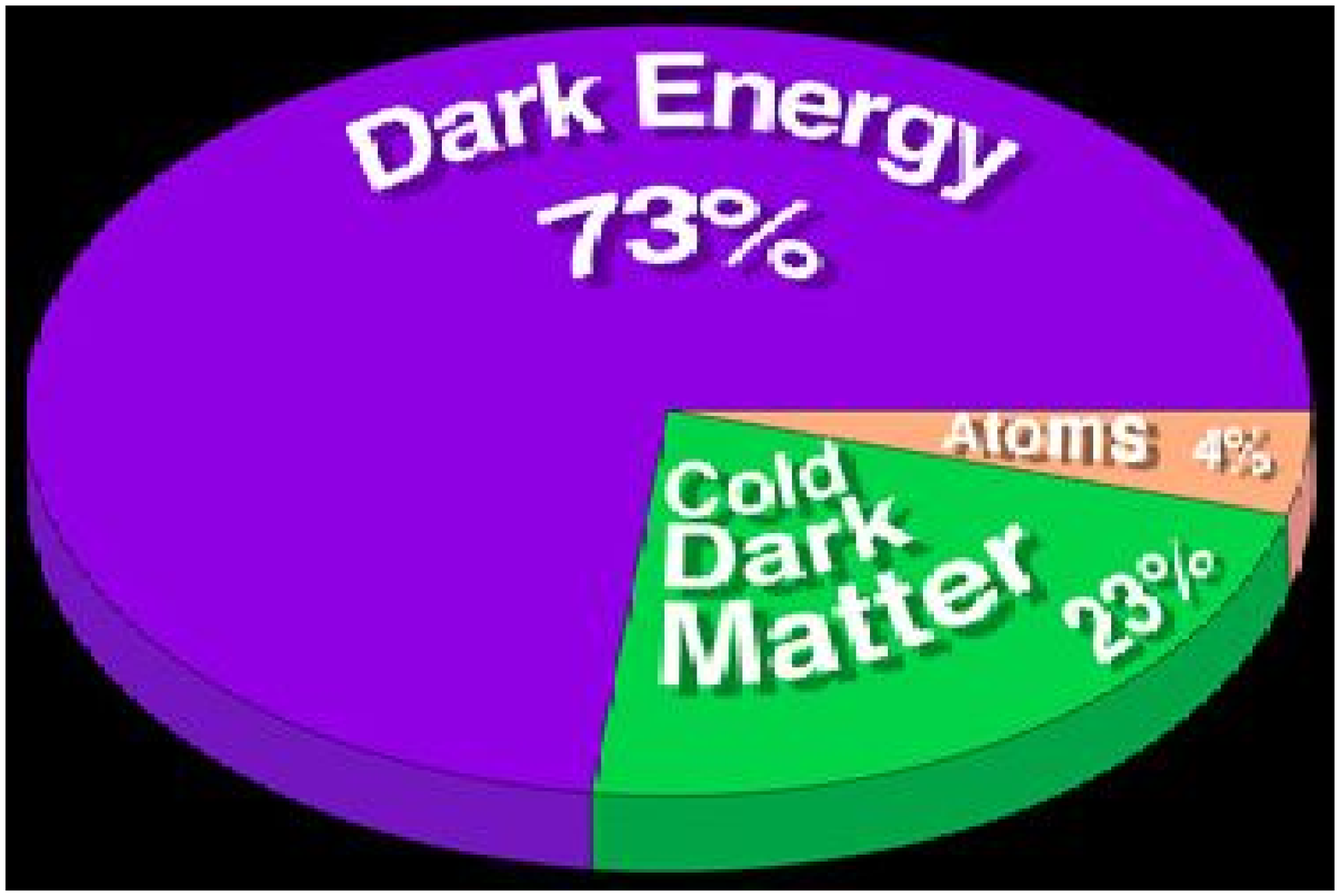}
\ffcaption{\label{fig:omega}
The energy content of the Universe according to WMAP data
(source: http://map.gsfc.nasa.gov/) }
\end{floatingfigure}

The evolution of the number density of supersymmetric relics in the
universe is described by the Boltzmann equation
as formulated for a 
FRW Universe. For calculations including 
many particle species, such as the case where co-annihilations
are important, there is a Boltzmann equation for each particle 
species~\cite{Griest:1990kh}, the equations can be combined to obtain
\begin{equation}
\frac{dn}{dt} =-3Hn-\langle\sigma_{eff}v\rangle\left(n^2-n_{eq}^2\right),\ 
\end{equation}
where
$n=\sum_{i=1}^{N} n_i,  \ \ \ 
n_{eq,i} =\frac{g_im_i^2T}{2\pi^2}K_2\left(\frac{m_i}{T}\right)$
and the sum extends over the $N$ particle species contributing to
the relic density, with $n_i$ being the number density of the $i$-th
species, $K_j$ is a modified Bessel function of the second kind of order $j$.
The quantity $\langle\sigma_{eff}v\rangle$ is the thermally averaged
cross section times velocity which is the key point in calculating relic density
including co-annihilation processes
with $\Omega h^2 \propto 1/[\int_0^{x_F}\langle \sigma_{eff}v\rangle dx]$,
where one integrates $\langle \sigma_{eff}v\rangle$ from 
0 to freeze-out temperature, 
$x_F$~\cite{Gondolo:1990dk,Griest:1990kh,Edsjo:1997bg}.
The main challenge in CDM calculation is the  evaluation
 of {\it all possible channels} for
neutralino annihilation  to SM and/or Higgs particles, as well as
all co-annihilation reactions
with {\it  relativistic thermal
averaging}. 
For our calculation of the neutralino relic density, we use the
IsaReD program~\cite{Baer:2002fv} interfaced with Isajet. 
There are also public programs for CDM calculations
such DarkSUSY~\cite{Gondolo:2000ee,Gondolo:2002tz}
and micrOMEGAs~\cite{Belanger:2001fz,Belanger:2004yn}
(as well as  many private ones)
which are in a good agreement with IsaReD.
\begin{figure}
\includegraphics[width=7.5cm]{%
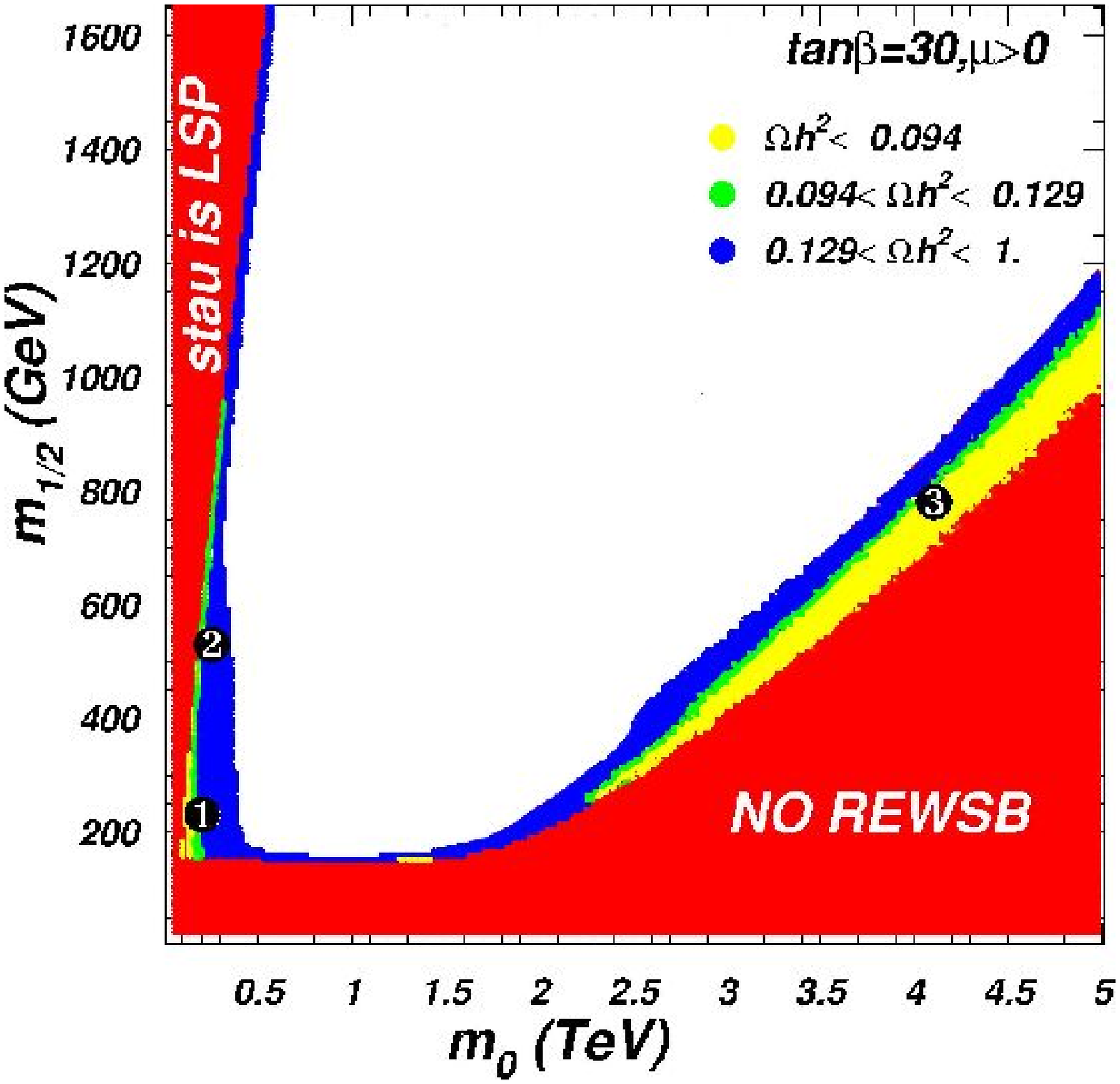}%
\includegraphics[width=7.5cm]{%
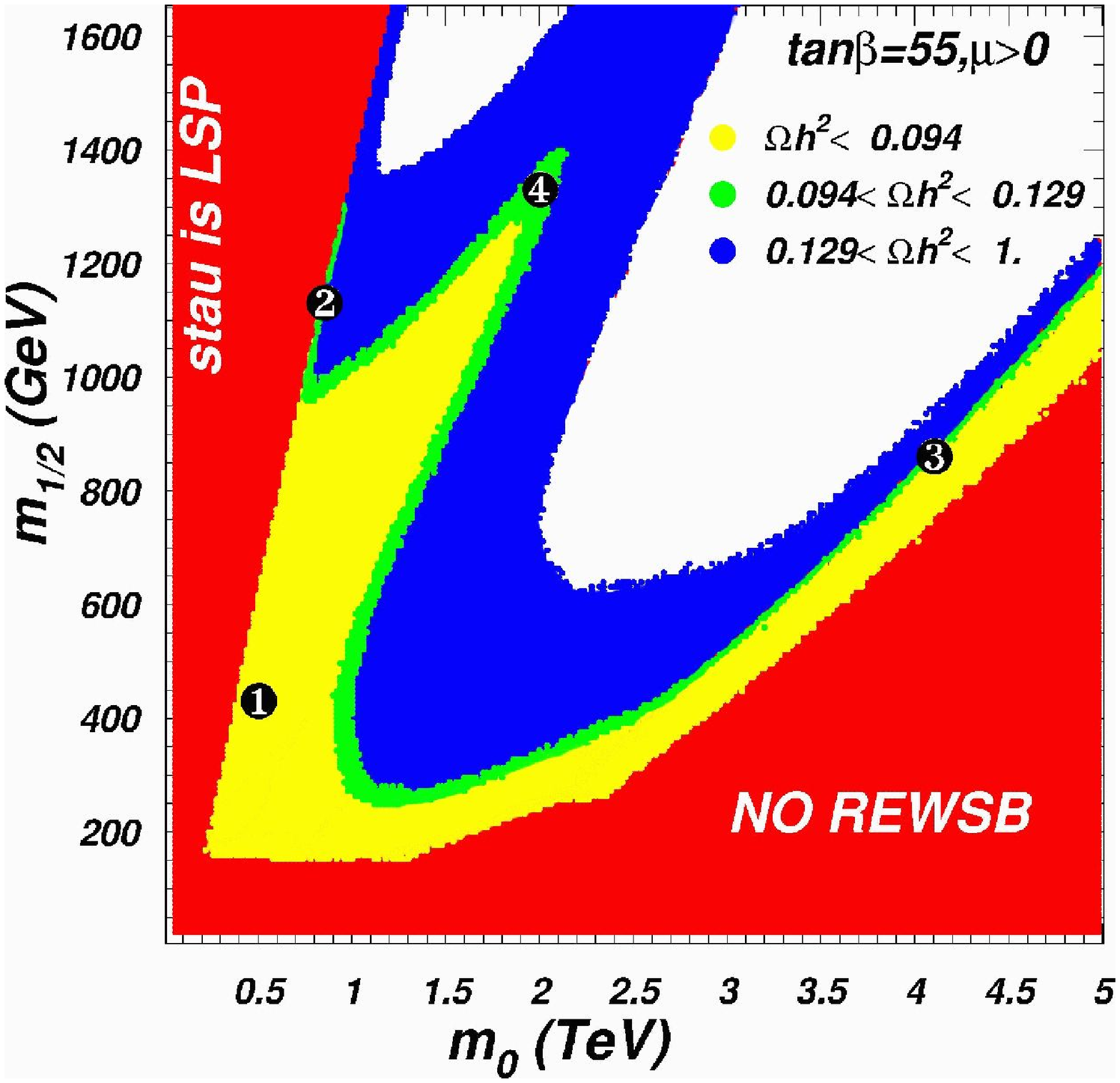}
\caption{\label{fig:relic} Regions of neutralino relic density in the $m_0$ vs 
$m_{1/2}$ plane for $A_0=0$ and $\tan\beta=30(55)$. Generic regions which are in agreement with WMAP results:
1)   bulk annihilation region, 2) stau co-annihilation region, 
3) hyperbolic branch/focus point (HB/FP) region and
4)  $A$-annihilation funnel region (only right plot for high $\tan\beta$).}
\end{figure}
In most of the parameter space of the mSUGRA model,  it turns out that a value
of $\Omega_{\tz_1}h^2$ well beyond the WMAP bound is generated. Only certain 
regions of the mSUGRA model parameter space give rise to a relatively low value
of $\Omega_{\tz_1}h^2$ in accord with WMAP measurements.  Fig.~\ref{fig:relic}
presents such generic regions in $m_0$ versus $m_{1/2}$ plane for $\mu>0$,
$A_0=0$  $\tan\beta=30, 55$ (left, right). The dark (red) shaded  regions are
excluded by theoretical constraints (lack of REWSB on the right, a charged LSP
in the upper left). The blue (dark-gray) shaded region
($0.129>\Omega_{\tz_1}h^2 >1$) and, furthermore, unshaded region 
($\Omega_{\tz_1}h^2 >1$)  are  excluded, since their too high relic density
would lead to overclosure of the Universe contradicting with WMAP data. The green
(gray) shaded regions are in exact 
agreement with WMAP constraints (Eq.~\ref{eq:wmap})
and could be classified in several classes:
\begin{enumerate}
\item The bulk annihilation region at low values of $m_0$ and $m_{1/2}$,
where neutralino pair annihilation occurs at a large rate via $t$-channel
slepton exchange.
\item The stau co-annihilation region at low $m_0$ where 
$m_{\tz_1}\simeq m_{\ttau_1}$ so that $\tz_1$s may co-annihilate
with $\ttau_1$s in the early universe~\cite{Ellis:1998kh,Ellis:2000we}.
\item The hyperbolic branch/focus point (HB/FP) 
region~\cite{Chan:1997bi,Feng:1999zg,Baer:1995va,Baer:1995nq} 
at large $m_0$
near the boundary of the REWSB excluded region where $|\mu |$ becomes
small, and the neutralinos have a significant higgsino component, 
which facilitates annihilations to 
$WW$ and $ZZ$ pairs.
\item The $A$-annihilation funnel, which occurs at very large 
$\tan\beta\sim 45-60$~\cite{Drees:1992am,Baer:1995nc,Baer:1997ai,%
Baer:2000jj,Ellis:2001ms,Roszkowski:2001sb}. 
In this case, the value of $m_A\sim 2m_{\tz_1}$.
An exact equality of the mass relation isn't necessary, since
the $A$ width can be quite large ($\Gamma_A\sim 10-50$ GeV);
then $2m_{\tz_1}$ can be several widths away from resonance,
and still achieve a large $\tz_1\tz_1\to A\to f\bar{f}$ annihilation
cross section. The heavy scalar Higgs $H$ also contributes to 
the annihilation cross section.  
\end{enumerate}

In addition, there exists a region of neutralino top-squark
co-annihilation (for very particular $A_0$ values) and a light Higgs
$h$ annihilation funnel (at low $m_{1/2}$ values).
In the light-gray (yellow) region neutralino relic density
is too low~($\Omega_{\tz_1}h^2 <0.096$) to give a good fit for
WMAP data. But, strictly speaking, this region is not excluded
since there could be {\it additional } particles (e.g. axions or particles from
various models with extra-dimensions) contributing to CDM relic density.
Therefore,  hereafter we treat green and yellow 
regions of~Fig.~\ref{fig:relic} as one, allowed ("green") region.

{\bf  Constraints from LEP2 searches}\\
The LEP2 collaborations have finished taking data and
collected the rich statistics
for 
$e^+e^-$ CM energies ranging up to $\sqrt{s}\simeq 208$ GeV.
Searches for superpartners at LEP2 gave negative results and led  to
the following constraints:
\begin{equation}
m_{\tw_1}>103.5~GeV  \mbox{~\cite{lep2-w1}}, 
\ \ \ \ 
m_{\te}>99~GeV \ \ \mbox{provided \ \ } m_{\tell}-m_{\tz_1}<10~GeV 
~\mbox{~\cite{lep2-sel}}
\end{equation}
The LEP2 experiments also searched for the SM Higgs boson. In addition to 
finding several compelling signal candidates consistent with 
$m_h\sim 115$ GeV, they set a limit
$m_{H_{SM}}>114.1$ GeV~\cite{unknown:2001xx}. In our mSUGRA parameter space scans, the
lightest SUSY Higgs boson $h$ is almost always SM-like. The exception occurs
when the value of $m_A$ becomes low, less than $100-150$ GeV.
This Higgs mass LEP2 limit requires radiative corrections
to the light Higgs mass (which is below Z-boson mass at the tree level)
to be quite large. Those radiative corrections  -- $\delta M_h$ --
are logarithmically sensitive to the SUSY scale ($M_{SUSY}$) and this
LEP2 limit pushes this  to level of $\sim 1~TeV$.
At the same time  $\delta M_h\propto m_{top}^4$. In the light of recent
$m_{top}$ measurements from D0~\cite{Abazov:2004cs} which shifted up
top-quark mass to be $178.0\pm 4.3$~GeV (as a world averaged value)
led to some relaxation of the lower limit on $M_{SUSY}$.

{\bf The $b\to s\gamma$ branching fraction}\\
The branching fraction $BF(b\to s\gamma )$ has recently been measured by the
BELLE~\cite{Abe:2001hk}, CLEO~\cite{Cronin-Hennessy:2001fk} and ALEPH~\cite{Barate:1998vz} collaborations.
Combining statistical and systematic errors in quadrature, these 
measurements give $(3.36\pm 0.67)\times 10^{-4}$ (BELLE), 
$(3.21\pm 0.43)\times 10^{-4}$ (CLEO) and $(3.11\pm 1.07)\times 10^{-4}$
(ALEPH). A weighted averaging of these results yields
$BF(b\to s\gamma )=(3.25\pm 0.37) \times 10^{-4}$. The 95\% CL range
corresponds to $\pm 2\sigma$ away from the mean. To this we should add 
uncertainty in the theoretical evaluation, which within the SM dominantly
comes from the scale uncertainty, and is about 10\%.\footnote{We caution
the reader that the SUSY contribution may have a larger theoretical
uncertainty, particularly if $\tan\beta$ is large.} Together, these 
imply the bounds,
\be
 2.2\times 10^{-4}< BF(b\to s\gamma )< 4.33 \times 10^{-4}
\label{eq:bsgamma}
\ee
Other computations of the range of $BF(b\to s\gamma )$ include for instance
Ellis {\it et al.}~\cite{Ellis:2002rp}: 
$2.33\times 10^{-4}<BF(b\to s\gamma )<4.15\times 10^{-4}$,
and Djouadi {\it et al.}~\cite{Djouadi:2001yk}: 
$2.0\times 10^{-4}<BF(b\to s\gamma )<5.0 \times 10^{-4}$.
In our study, we simply show contours of $BF(b\to s\gamma )$ of
2, 3, 4 and $5\times 10^{-4}$, allowing the reader the flexibility
of their own interpretation.

The calculation of $BF(b\to s\gamma )$ used here is based upon the
program of Ref.~~\cite{Baer:1996kv,Baer:1997jq}. That calculation uses an effective field 
theory approach to evaluating radiative corrections to the 
$b\to s\gamma$ decay rate. 
In our calculations, we implement the running $b$-quark mass
including SUSY threshold corrections as calculated in ISASUGRA;
these effects can be important at large values of the 
parameter $\tan\beta$~\cite{Degrassi:2000qf,Carena:2000uj}.
Once the relevant operators and Wilson coefficients are 
known at $Q=M_W$, then the SM WCs are evolved down to $Q=m_b$ via
NLO RG running. At $m_b$, the $BF(b\to s\gamma )$ is evaluated at NLO,
including bremsstrahlung effects. Our value of the SM $b\to s\gamma$
branching fraction yields $3.4\times 10^{-4}$, with a scale uncertainty
of 10\%.     

{\bf Muon anomalous magnetic moment}\\
The muon anomalous magnetic moment $a_\mu =\frac{(g-2)_\mu}{2}$ has been
recently measured to high precision by the E821 experiment~\cite{Bennett:2004pv}
for
the negative muon along with earlier results on the positive muon~\cite{Bennett:2002jb}.
In addition, theoretical determinations of $(g-2)_\mu$ have been
presented by Davier {\it et al.}~\cite{Davier:2003pw} and 
Hagiwara {\it et al.}~\cite{Hagiwara:2003da} which use recent data on 
$e^+e^-\to hadrons$ at low energy
to determine the hadronic vacuum polarization contribution to the
muon magnetic moment. Combining the latest experiment and theory
numbers, we find the deviation of $a_\mu$ to be:
\begin{eqnarray}
\Delta a_\mu &=&(27.1\pm 9.4)\times 10^{-10}\ \ \ ({\rm Davier}\ et\ al.)\\
\Delta a_\mu &=&(24.7\pm 9.0)\times 10^{-10}\ \ \ ({\rm Hagiwara}\ et\ al.) .
\label{eq:amu}
\end{eqnarray}

The Davier {\it et al.} group also presents a number using $\tau$ decay data to
determine the hadronic vacuum polarization, which gives
$\Delta a_\mu =(12.4\pm 8.3)\times 10^{-10}$, {\it i.e.} nearly consistent
with the SM prediction. However, there seems to be growing consensus
that the numbers using the $e^+e^-$ data are to be trusted more, since 
they offer a direct determination of the hadronic vacuum polarization.
The $\sim 3\sigma$ deviation in $a_\mu$ using the $e^+e^-$ data 
can be explained in a supersymmetric
context if second generation sleptons (smuons and muon sneutrinos)
and charginos and neutralinos are relatively light.

{\bf $B_s\to\mu^+\mu^-$ decay}\\
While all SUSY models contain two doublets of Higgs superfields, there
are no tree-level flavor changing neutral currents because one doublet
${\hat H}_u$ couples only to $T_3=1/2$ fermions, while the other doublet
$\hat{H}_d$ couples just to $T_3= -1/2$ fermions. At one loop, however,
couplings of ${\hat H}_u$ to down type fermions are induced. These
induced couplings grow with $\tan\beta$. As a result, down quark Yukawa
interactions and down type quark mass matrices are no longer
diagonalized by the same transformation, and flavor violating couplings
of neutral Higgs scalars $h$, $H$ and $A$ emerge. Of course, in the limit of
large $m_A$, the Higgs sector becomes equivalent to the SM
Higgs sector with the light Higgs boson $h=H_{SM}$, and the flavor
violation decouples. The interesting thing is that while
this decoupling occurs as $m_A \to \infty$, {\it there is no decoupling
for sparticle masses becoming large.}

An important consequence of this coupling is the possibility of the
decay $B_s \to \mu^+\mu^-$, whose branching fraction has been
experimentally
bounded by CDF~\cite{Abe:1998ah} to be:
\be
 BF(B_s\to\mu^+\mu^- )< 2.6\times 10^{-6}, 
\ee
mediated by
the neutral states in the Higgs sector of supersymmetric models. While
this branching fraction is very small within the SM ($BF_{SM}(B_s \to
\mu^+\mu^-)\simeq 3.4 \times 10^{-9}$), 
the amplitude for the Higgs-mediated decay
of $B_s$ grows as $\tan^3\beta$ within the SUSY framework, and hence can
completely dominate the SM contribution if $\tan\beta$ is large.
Several groups~\cite{Arnowitt:2002cq,Dedes:2001fv,Babu:1999hn} 
and recently~\cite{Mizukoshi:2002gs}
have analyzed the implications of this decay
within the mSUGRA framework.
We also present improved $b\to s\gamma$ branching
fraction predictions in accord with~\cite{Mizukoshi:2002gs} for the current ISAJET release. This
constraint is important only for very large values of
$\tan\beta$.

{\bf mSUGRA constraints.}\\
\begin{ltxfigure}
\epsfig{width=7.cm,file=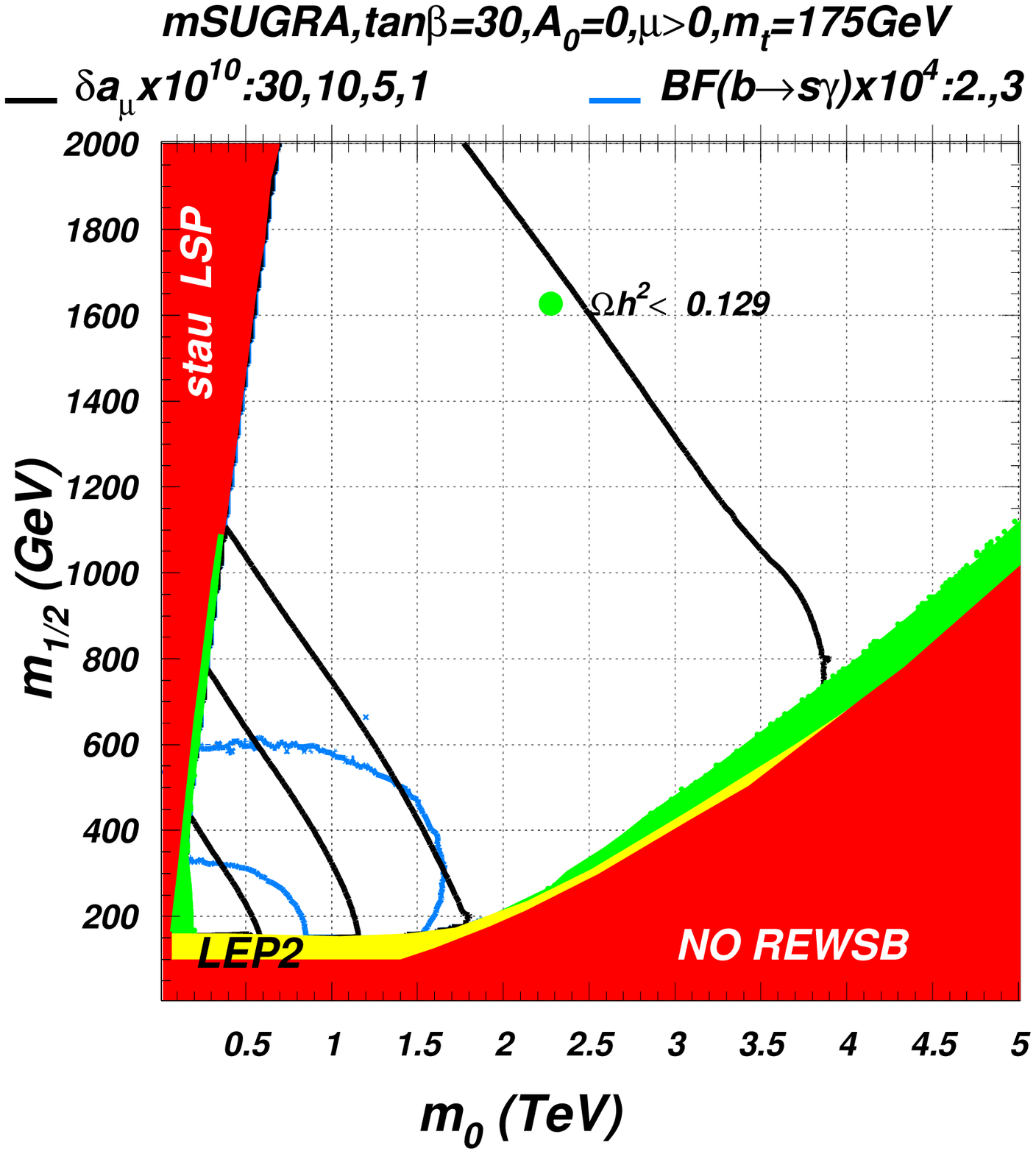}
\epsfig{width=7.cm,file=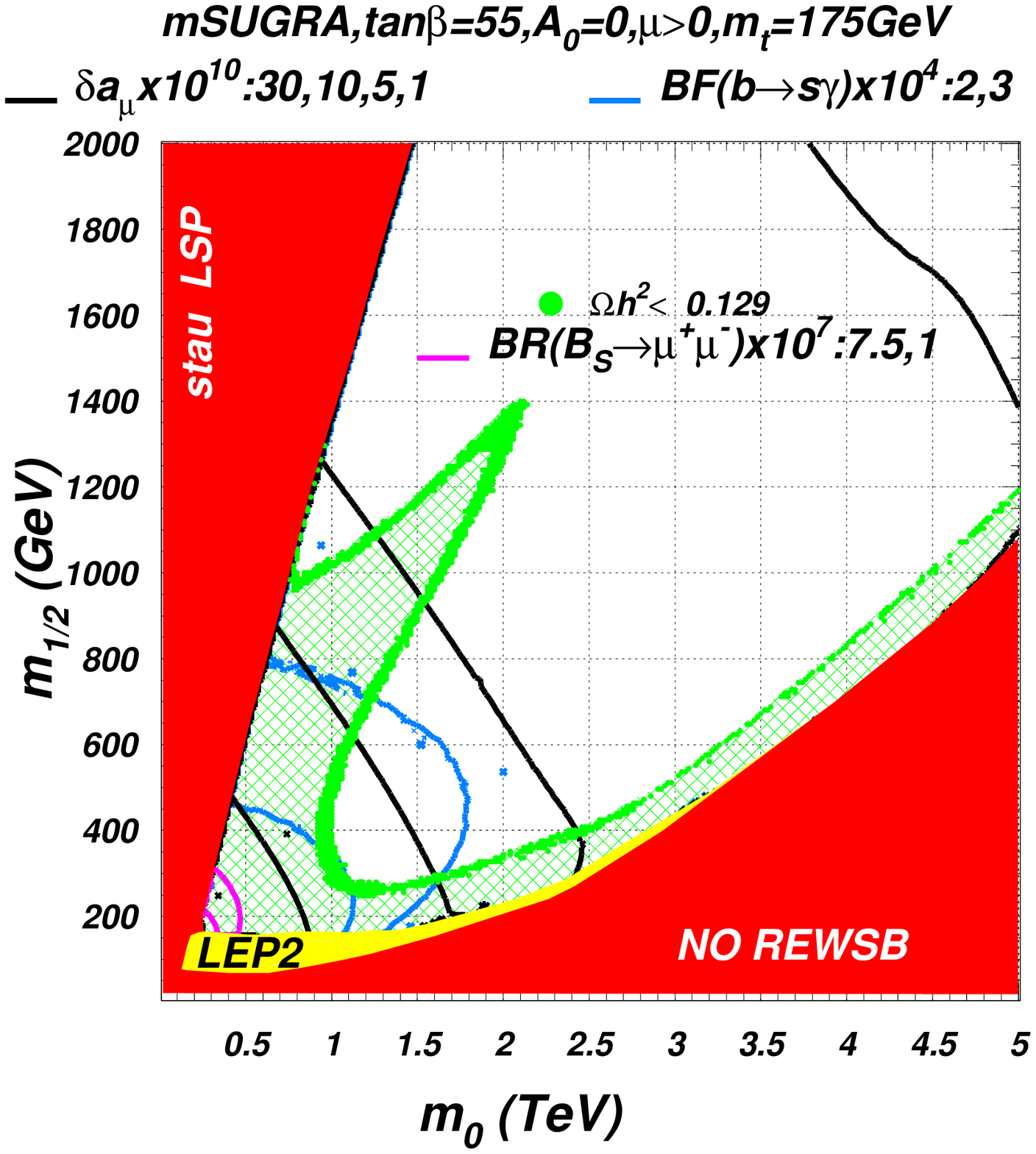}\\
\epsfig{width=7.cm,file=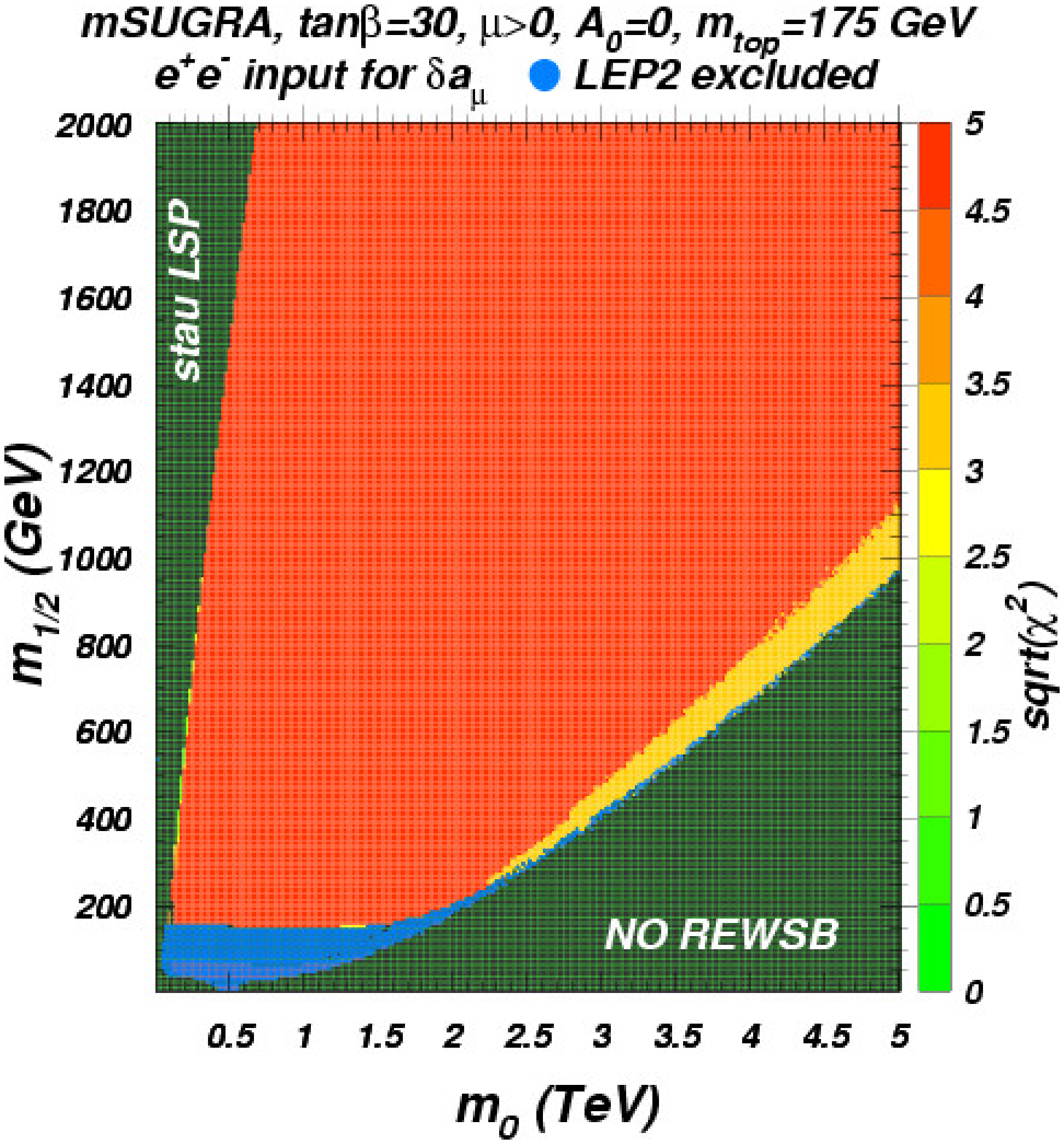}
\epsfig{width=7.cm,file=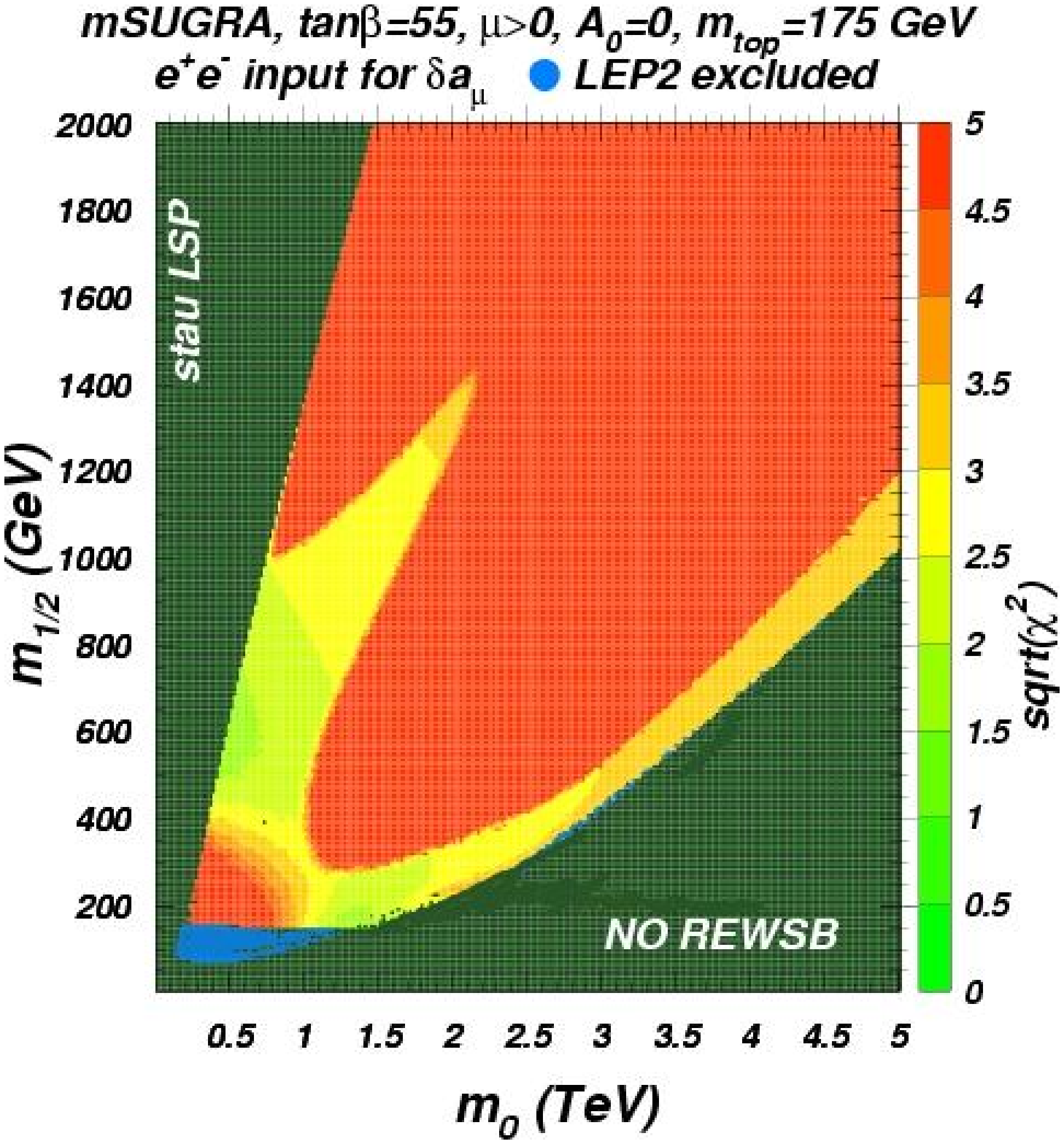}
\caption{ Constraints for
mSUGRA model in the $m_0$ vs 
$m_{1/2}$ plane for $A_0=0$ and $\tan\beta=30(55)$ (top raw).
Top raw:
allowed CDM relic density regions are the green shaded ones,
LEP2 excluded regions denoted by yellow color.
Black contours denotes $\delta a_\mu$ 
values ($30,10,5,1\times 10^{-10}$ -- from bottom to top)
and blue contours denote $BF(b\to s\gamma)$ 
values ($2,3\times 10^{-4}$ -- from bottom to top).
Magenta contours for $BF(B_S\to \mu^+\mu^-)$ are relevant only for 
very high $\tan\beta=55$ ($7.5,1\times 10^{-7}$ -- from bottom to top).
The bottom raw of figures presents $\sqrt{\chi^2}$ 
for the respective parameters of the mSUGRA model.
The green regions have low $\chi^2$, while red regions have
high $\chi^2$. Yellow is intermediate.}
\label{fig:relic-amu}
\end{ltxfigure}
Within the mSUGRA framework, the parameters $m_0$ and $m_{1/2}$ are the
most important for fixing the scale of sparticle masses. The
$m_0-m_{1/2}$ plane (for fixed values of other parameters) is convenient
for
a simultaneous display of these constraints, and hence, of parameter
regions in accord with all experimental data.
Physicists interested in the mSUGRA model may 
wish to focus their attention on these regions.

Our   results for combined constraints mentioned above are 
presented in Fig.~\ref{fig:relic-amu} for mSUGRA model in the $m0$ vs 
$m_{1/2}$ plane for $A_0=0$ and $\tan\beta=$30~(55) in the left (right)
frame, respectively.  
Top raw of frames presents constraints themselves, while the bottom raw 
presents corresponding $\chi^2$ pattern for frames above.
$\chi^2$ is formed from
$\Delta a_\mu$, $BF(b\to s\gamma )$ and $\Omega_{\tz_1}h^2$, as given in 
Eqn's (\ref{eq:wmap},\ref{eq:bsgamma},\ref{eq:amu}).

Allowed CDM relic density regions are the green shaded ones,
LEP2 excluded regions denoted by yellow color.
Black contours denotes $\delta a_\mu$ 
values ($30,10,5,1\times 10^{-10}$ -- from bottom to top)
and blue contours denote $BF(b\to s\gamma)$ 
values ($2,3\times 10^{-4}$ -- from bottom to top).
In the  left upper frame of Fig.~\ref{fig:relic-amu}
one can observe  the  tension
between  $\delta a_\mu$ and $BF(B\to s\gamma)$.
The point is that while $BF(b\to s\gamma )$ favors large
third generation squark masses to suppress 
SUSY contributions to $b\to s\gamma $
decay, $\Delta a_\mu$ experimental value favors relatively light second generation
slepton masses, to give a significant  $(g-2)_\mu$ deviation  from
the SM value. 
This is clearly reflected  in the respective (left bottom) frame
where the $\chi^2$ quantity is presented.

In case of low and intermediate values of $\tan\beta$
almost all the 
$m_0\ vs.\ m_{1/2}$ plane has very large $\chi^2$. This arises because 
in general an overabundance of dark matter is produced in the 
early universe, 
and the relic density $\Omega_{\tz_1}h^2$ is beyond WMAP limits.
There is a very narrow sliver of yellow at $m_{1/2}\sim 150$ GeV
(just beyond the LEP2 limit) where $2m_{\tz_1}\simeq m_h$, and neutralinos
can annihilate through the narrow light Higgs resonance. 
In addition, there is an orange/yellow region at high $m_0$ 
at the edge of parameter space (the HB/FP region), with an intermediate
value of $\chi^2$. In an earlier study~\cite{Baer:2003yh}, 
this region was found to have a low
$\chi^2$ value. In the present situation, however, the $3\sigma$ deviation from the SM of
$a_\mu$ tends to disfavor the HB/FP region. In the HB/FP region, 
sleptons are so heavy (typically 3-5 TeV), that SUSY contributions to 
$a_\mu$ are tiny, and the prediction is that $a_\mu$ should be in near 
accord with the SM calculation. The remaining green region is the narrow sliver
that constitutes the stau co-annihilation region, 
barely visible at the left hand edge of parameter space 
adjacent to where $\ttau_1$ becomes the LSP.

Once we move to very large $\tan\beta$ values,
as shown in the right set of frames of Fig.~\ref{fig:relic-amu}, 
then the $A$-annihilation funnel
becomes visible, and some large regions of moderately low $\chi^2$ 
appear around $m_0,\ m_{1/2}\sim 500,\ 600$ GeV and also at
$1500,\ 200$ GeV. While the $A$-annihilation funnel extends
over a broad region of parameter space, the upper and lower ends of
the funnel are disfavored: basically, if sparticles become too 
heavy (the upper end), then
$\Delta a_\mu$ becomes too small, 
while if sparticles become too light (the lower end), 
then $BF (b\to s\gamma )$
deviates too much from its central value.

One comes to conclusions~\cite{Baer:2004xx} that  for the mSUGRA model almost all of
parameter space is excluded or at least disfavored by the combination
of the WMAP $\Omega_{\tz_1}h^2$ limit, the new $\Delta a_\mu$
value, and the $BF(b\to s\gamma )$ value. The $\Omega_{\tz_1}h^2$ 
constraint only allows the several regions of parameter space mentioned 
above, while $BF(b\to s\gamma )$ favors large
third generation squark masses to suppress 
SUSY contributions to $b\to s\gamma $
decay, and $\Delta a_\mu$ favors relatively light second generation
slepton masses, to give a significant deviation of $(g-2)_\mu$ from
the SM value. The only surviving regions with relatively low
$\sqrt{\chi^2}\alt 2$ are the stau co-annihilation region, and 
intermediate portions of the $A$-annihilation funnel at very large
values of $\tan\beta$.
One should note, however,  that if the hadronic vacuum polarization determination 
using $\tau$ decay data turn out to be correct, then the
HB/FP region will appear in a more favorable light!
%

{\bf Reach of LHC and LC in Dark Matter Allowed Regions of the mSUGRA Model}
\\
The reach of the CERN LHC for SUSY in the mSUGRA model has been 
calculated in Ref.~\cite{Baer:2003wx} assuming 100 fb$^{-1}$ of
integrated luminosity. Briefly, sparticle pair production events
were generated for many mSUGRA model parameter choices in the 
$m_0\ vs.\ m_{1/2}$ plane for various $\tan\beta$ values. 
A fast LHC detector simulation (CMSJET) was used, and cuts were imposed to
extract signal rates in a variety of multilepton plus multijet plus
missing transverse energy channels. Backgrounds were also calculated
from a variety of QCD and vector boson production processes. A large
set of selection cuts were used to give some optimization over 
broad regions of parameter space. It was required to have at least a
$5\sigma$ signal over background, with at least 10 signal events 
in the sample.

\begin{figure}
\epsfxsize=7.5cm\epsffile{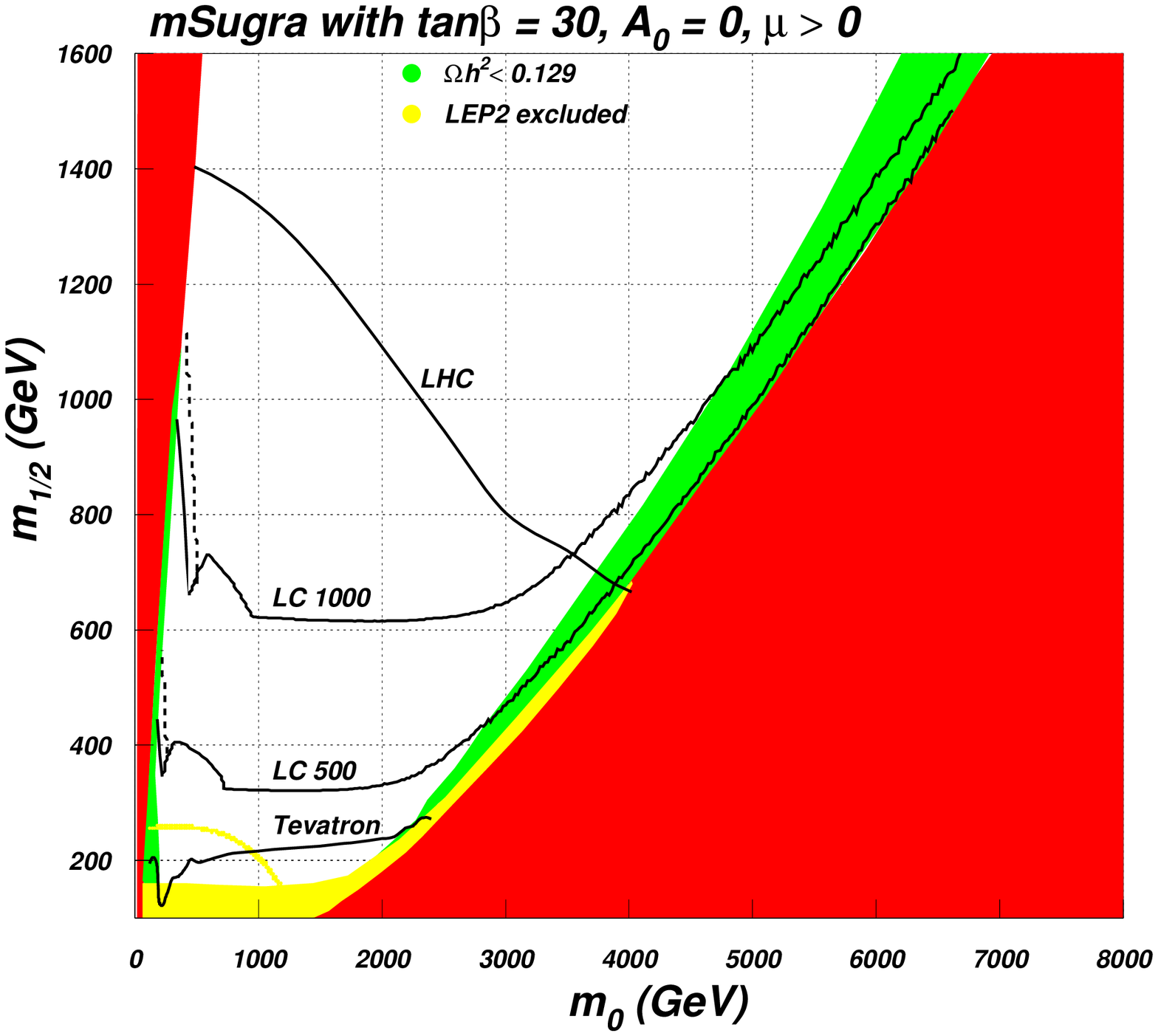} 
\epsfxsize=7.5cm\epsffile{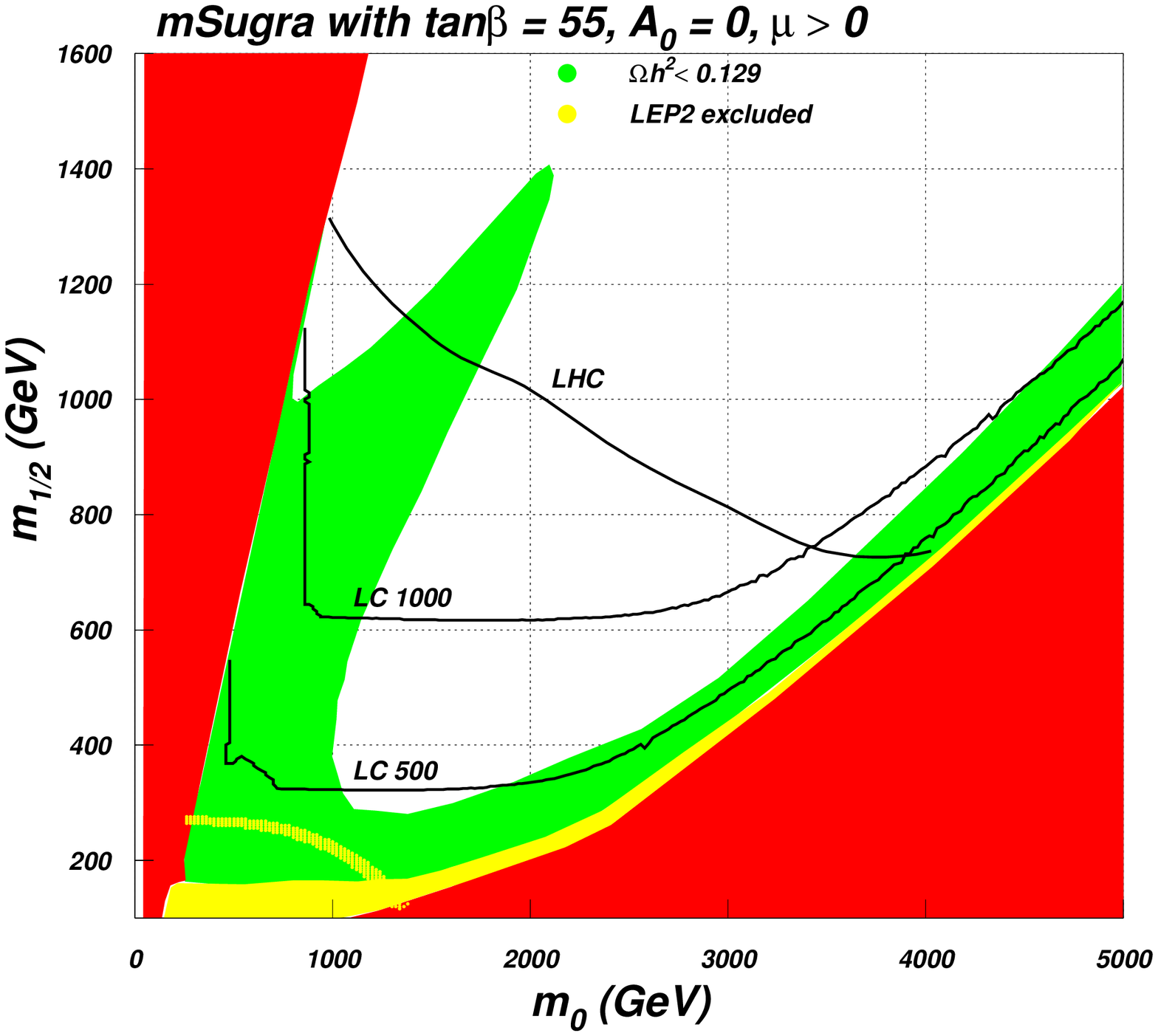}
\caption{ \label{fig:lhcreach}Parameter space of mSUGRA model for $\tan\beta =30(55)$left(right),
$A_0=0$ and $\mu >0$, showing the reach of the Fermilab Tevatron, 
the CERN LHC 
and a 0.5 and 1 TeV linear $e^+e^-$ collider for supersymmetry discovery.} 
\end{figure}

The reach of the CERN LHC is shown in Fig. \ref{fig:lhcreach}
for the case of $\tan\beta =30~(55)$ left (right), $\mu >0$, $A_0=0$ and $m_t=175$ GeV.
The dark shaded (red) regions are disallowed by lack of 
radiative electroweak symmetry
breaking (REWSB) (right hand side) or presence of a stau LSP (left hand side).
The light gray (yellow) region is excluded by LEP2 chargino searches 
($m_{\tilde{\chi}^+}>103.5$ GeV), while the region below the yellow 
contour gives $m_h<114.4$ GeV, in contradiction of LEP2 SM Higgs searches
(here, the SUSY $h$ Higgs boson is essentially SM-like). The 
medium gray (green) regions
have $\Omega_{CDM}h^2<0.129$, and are {\it allowed} by WMAP. The broad
HB/FP region is seen on the right-hand side, while the stau co-annihilation 
region is shown on the left-hand side. At the edge of the LEP2 excluded 
region is the light Higgs annihilation corridor. The reach of the 
Fermilab Tevatron via the trilepton channel is also shown~\cite{Baer:2003dj}, 
assuming a $5\sigma$ signal over background for 10 fb$^{-1}$. 
The reach of the CERN LHC
for 100 fb$^{-1}$ of integrated luminosity is shown by the contour labeled
``LHC''. It extends from $m_{1/2}\sim 1400$ GeV (corresponding to
a value of $m_{\tilde g}\sim 3$ TeV) on the left-hand side, to
$m_{1/2}\sim 700$ GeV (corresponding to $m_{\tilde g}\sim 1.8$ TeV) 
on the right-hand side. In particular, for those values of $\tan\beta$, the
LHC reach covers the entire stau co-annihilation region, plus the
low $m_{1/2}$ portion of the HB/FP region. The outer limit of the reach
contour is mainly determined by events in the $E_T^{miss}+$ jets channel,
which arises from gluino and squark pair production, followed by hadronic 
cascade decays.

We also show in the plot the reach of a $\sqrt{s}=500$ and 1000 GeV
LC, assuming 100 fb$^{-1}$ of integrated luminosity~\cite{Baer:2003ru}. 
Events were generated
using Isajet 7.69, and compared against various SM backgrounds.
The left-most portion of the reach contour arises where selectron and smuon
pair production are visible, while the main portion (flat with $m_{1/2}$)
arises due to chargino pair searches. An additional reach is gained between
these two regions by searching for $e^+e^-\rightarrow \tilde{\chi}_2^0
\tilde{\chi}_1^0$ production, followed by 
$\tilde{\chi}_2^0\rightarrow \tilde{\chi}_1^0 b\bar{b}$ decay. 
In addition, in Ref.~\cite{Baer:2004zk}, additional reach can be gained by 
searching for stau pair events, 
although two photon backgrounds must be accounted for, 
due to the low energy release in the stau co-annihilation region.

\begin{floatingfigure}{7cm}
\hspace*{-1cm}\epsfxsize=8cm\epsffile{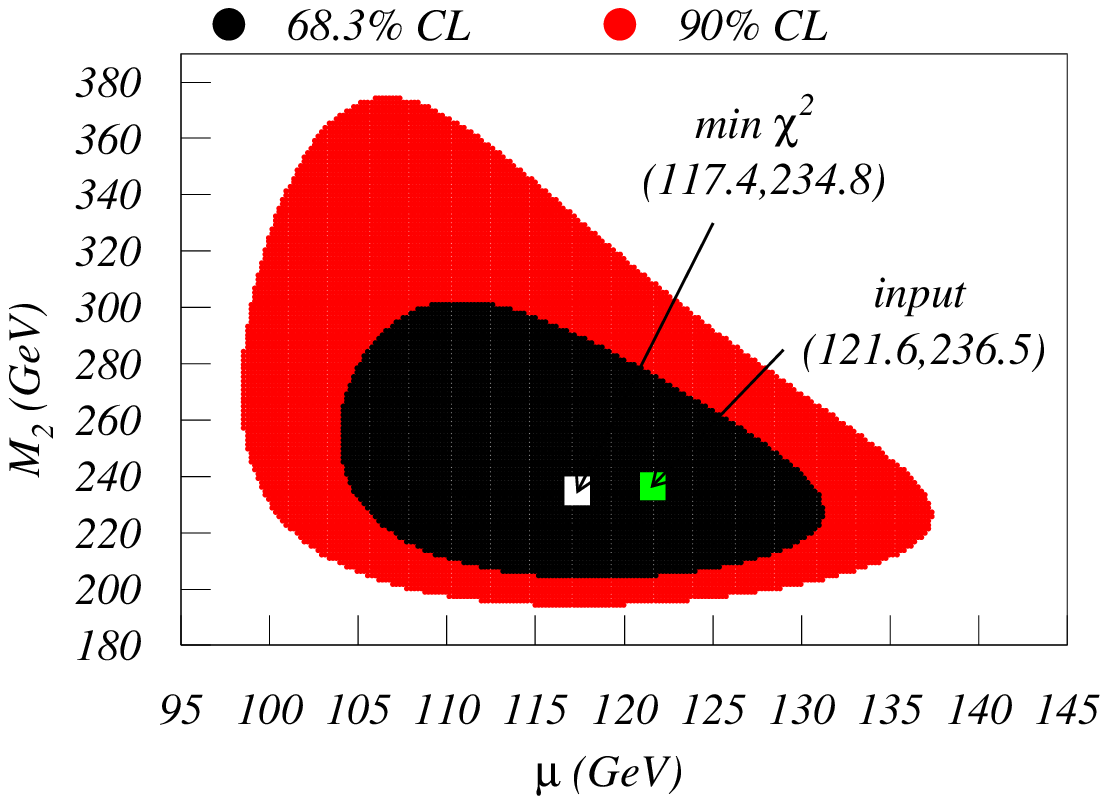}
\ffcaption{ \label{sec442_fig3}
Determination of SUSY parameters from 
examining chargino pair production at a $\sqrt{s}=0.5$ TeV LC,
for the HB/FP mSUGRA point listed in the text.}
\end{floatingfigure}

While a 500 GeV LC can cover only a portion of the stau co-annihilation
region, a 1 TeV LC can cover the entire region, at least for this value
of $\tan\beta$. As one moves into the HB/FP region, the LC retains
a significant reach for SUSY, which in fact extends {\it beyond} that
of the CERN LHC! It is significant that this additional reach occurs in a 
DM allowed region of parameter space. In the HB/FP region, the
superpotential $\mu$ parameter becomes small, and the lightest chargino 
and neutralino become increasingly light, with increased higgsino content. 
In fact, the 
decreasing mass gap between $\tilde{\chi}_1^+$ and $\tilde{\chi}_1^0$
makes chargino pair searches difficult at a LC using conventional cuts
because there is so little visible energy release from the chargino
decays. In Ref.~\cite{Baer:2003ru,Baer:2004zk}, 
we advocated cuts that pick out low energy release signal events
from SM background, and allow a LC reach for chargino pairs 
essentially up to the kinematic limit for their production.
In this case, it is important to fully account for $\gamma\gamma\to f\bar{f}$
backgrounds, where $f$ is a SM fermion.

The case study of Ref.~\cite{Baer:2003ru} also examined the HB/FP region
with parameters $m_0=2500$ GeV, $m_{1/2}=300$ GeV, $A_0=0$, $\tan\beta = 30$,
$\mu >0$ and $m_t=175$ GeV, {\it i.e.} in the HB/FP region. 
In this case, chargino pair events were selected from the 
$1\ell+$jets $+E_T^{miss}$ channel, and the dijet mass distribution was used
to extract the value of $m_{\tilde{\chi}_1^+}$ and $m_{\tilde{\chi}_1^0}$
at the 10\% level. The mass resolution is somewhat worse than 
in previous case studies in the literature because the charginos undergo
three-body rather than two-body decays, and no sharp edges in energy
distributions are possible. Nonetheless, the measured value of
chargino and neutralino mass, along with a measure of the total
chargino pair
cross section, was enough to determine the SUSY parameters $M_2$ and $\mu$
to 10-20\% precision. The results, shown in Fig.~8, 
demonstrate that $\mu <M_2$, which points to a $\tilde{\chi}_1^+$ and
$\tilde{\chi}_1^0$ which are higgsino/gaugino mixtures, as is
characteristic of the HB/FP region. 

In conclusion to this section,
one should stress that the CERN LHC and an $e^+e^-$
LC are highly complementary to each other in
exploring the dark matter allowed parameter space of the mSUGRA model.
LHC covers the  
stau co-annihilation region (completely for $\tan\beta <40$)
as well as the $H,\ A$ funnel region
(much of which is  typically beyond 
the maximum reach of a LC).
However only the lower part  of the HB/FP region can be covered by the LHC.
On the other hand, as we have demonstrated, LCs 
can probe much of the {\it upper} part of the HB/FP region
with the new proposed cuts.
Therefore, the combination of the LHC and a TeV scale 
LC can cover almost the entire  parameter space of the mSUGRA scenario.

{\bf Direct and Indirect CDM searches}\\
Besides collider experiments there exist both direct
and indirect non-accelerator  dark matter search  experiments,
which are ongoing and proposed. Those experiments
as will be discussed below play crucial role in restricting
SUSY parameter space.
\\
{\bf Direct CDM } search experiments are aimed to look  at neutral dark matter candidates
scattering off nuclei. 
Direct dark matter detection has been recently 
examined by many authors~\cite{Baer:2003jb}, and observable signal rates
are generally found in either the bulk annihilation region, or in the
HB/FP region, while direct detection of DM seems unlikely in the 
$A$-funnel or in the stau co-annihilation region.
Early
limits on the spin-independent neutralino-nucleon cross-section 
($\sigma_{SI}$) have 
been obtained by the CDMS~\cite{Abrams:2002nb}, EDELWEISS~\cite{Benoit:2002hf} 
and ZEPLIN1~\cite{Spooner:2001zx} groups, while a signal was 
claimed by the DAMA collaboration~\cite{Bernabei:2003xg}. 
Collectively, we will refer to the reach from these groups as the ``Stage 1''
dark matter search. Depending on the neutralino mass,
the combined limit on $\sigma_{SI}$
varies from  $10^{-5}$ to $10^{-6}$~pb. This cross section
range is beyond the predicted levels from most supersymmetric models.
However, experiments in the near future
like CDMS2, CRESST2~\cite{Bravin:1999fc}, ZEPLIN2 and EDELWEISS2 
(Stage 2 detectors) should have a 
reach of the order of 
$10^{-8}$~pb.
In fact, the first results from CDMS2 have recently appeared, and yield
a considerable improvement over the above mentioned Stage 1 
results~\cite{Akerib:2004fq}.
Finally, a number of experiments such as GENIUS~\cite{Klapdor-Kleingrothaus:2003pe}, 
ZEPLIN4~\cite{Cline:2003pi} and XENON~\cite{Suzuki:2000ch} are in 
the planning stage. We refer to these as Stage 3 detectors, which promise 
impressive limits of the order of $\sigma_{SI}<10^{-9}$ -- $10^{-10}$~pb, and 
would allow the exploration of a considerable part of parameter space of many 
supersymmetric models. In particular, the Stage 3 direct DM detectors
should be able to probe almost the entire HB/FP region of mSUGRA model 
parameter space. We note here in addition 
that the Warm Argon Program (WARP)~\cite{Brunetti:2004cf}
promotes a goal of detecting neutralino-nucleus scattering cross sections 
as low $10^{-11}$ pb.
\\
Indirect detection of neutralino dark matter may occur via
(for review, see {\it e.g.}~\cite{Eigen:2001mk,deBoer:2003jt,Hooper:2003ka}
\begin{enumerate}
\item observation of high energy neutrinos originating from
$\tz_1\tz_1$ annihilations in the core of the sun or earth
\item observation of $\gamma$-rays originating from neutralino annihilation
in the galactic core or halo 
and 
\item observation of positrons or anti-protons
originating from neutralino annihilation in the galactic halo. 
\end{enumerate}

The latter
signals would typically be non-directional due to the influence of galactic
magnetic fields, unless the neutralino annihilations occur relatively
close to earth in regions of clumpy dark matter.
The indirect signals for SUSY dark matter have been investigated 
in a large number of papers, and computer codes which yield
the various signal rates are available~\cite{Jungman:1995df,Gondolo:2002tz}. 
Recent works find that the various indirect signals occur at 
large rates in the now disfavored bulk annihilation region, 
and also in the HB/FP region~\cite{Feng:2000gh,Feng:2000zu}.
In Ref.~\cite{Baer:2003bp}, it was pointed out that the $A$ annihilation
funnel can give rise to large rates for cosmic $\gamma$s, $e^+$s and
$\bar{p}$s. However, neutralino-nucleon scattering cross sections 
are low in the $A$ annihilation funnel, so that 
no signal is expected at neutrino telescopes, which depend more on the
neutralino-nucleus scattering cross section than on the neutralino
annihilation rates.

There is a bunch of various experiments for indirect dark matter detection.
{\bf Neutrino telescopes} such as Antares or IceCube
are aimed to detect muon neutrinos from neutralino annihilation
in the core of the earth or sun via $\nu_\mu\to \mu$ conversions.
The Antares $\nu$ telescope
should be sensitive to $E_\mu >10$ GeV; it is in the process of deployment 
and is expected to turn on in 2006~\cite{Carmona:2001ye}.
It should attain a sensitivity of $100-1000$ $\mu$s/km$^2$/yr. 
The IceCube $\nu$ telescope is also in the
process of deployment at the south pole~\cite{Ahrens:2002dv,Halzen:2003fi}. 
It should be sensitive to
$E_\mu >25-50$ GeV, and is expected to attain a sensitivity of 
$10-100$ $\mu$s/km$^2$/yr. Full deployment of all detector elements is
expected to be completed by 2010.

Neutralino could also annihilate
in the galactic halo giving rise {\bf gamma rays, positrons or anti-protons}
fluxes. 
The $\gamma$ rays 
can be detected down to sub-GeV energies with space-based
detectors such as EGRET~\cite{Mayer-Hasselwander:1998hg} or GLAST~\cite{Morselli:2002nw}. 
Ground based arrays require much higher
photon energy thresholds of order $20-100$ GeV. Experiments such as 
GLAST should be sensitive to rates of order $10^{-10}$ $\gamma$s/cm$^2$/sec
assuming $E_\gamma >1$ GeV. 
In fact, it has recently been suggested that the extra-galactic gamma ray 
background radiation as measured by EGRET is well fit by a model
of neutralino annihilation~\cite{Elsaesser:2004ap,deBoer:2003ky}. In~\cite{deBoer:2004ab}
detailed study has been done explaining not only the rate of SUSY contribution to the 
EGRET data but also the shape of various distribution through the non-trivial shape 
of CDM  halo.

Positrons would arise as decay products of
heavy quarks, leptons and gauge bosons produced in neutralino annihilations.
Space based anti-matter detectors such as Pamela~\cite{Pearce:2002ef} 
and AMS-02~\cite{Casaus:2003xt} will be able 
to search for anomalous positron production from dark matter annihilation.
The cosmic positron excess as measured by HEAT~\cite{DuVernois:2001bb} 
has been suggested as 
having a source in galactic halo neutralino annihilations~\cite{Baltz:2001ir,deBoer:2003ky}.
It is suggested by Feng {\it et al.}~\cite{Feng:2000gh,Feng:2000zu} 
that a reasonable observability criteria
is that 
signal-to-background ($S/B$) rates 
should be greater than the 1-2\% level.
To calculate the
$S/B$ rates, we adopt fit C from Ref.~\cite{Feng:2000gh,Feng:2000zu} for the  
$E^2 d\Phi_{e^+}/d\Omega dE$ background rate:
\be
E^2 d\Phi_{e^+}/d\Omega dE =1.6\times 10^{-3}\ E^{-1.23},
\ee
where $E$ is in GeV.

Anti-protons may also be produced in the debris of neutralino annihilations 
in the galactic halo. Such anti-protons have been measured by the 
BESS collaboration~\cite{Orito:1999re}.
The differential flux of anti-protons from
the galactic halo, $d\Phi_{\bar{p}}/dE_{\bar{p}}d\Omega$, 
as measured by BESS, has a peak in the kinetic energy distribution 
at $E_{\bar{p}}\sim 1.76$ GeV.
The height of the peak
at $E_{\bar{p}}\sim 1.76$ GeV is $\sim 2\times 10^{-6}$ 
${\bar{p}}$/GeV/cm$^2$/s/sr. Signal rates in the range
of $10^{-7}-10^{-6}$ ${\bar{p}}$/GeV/cm$^2$/s/sr might thus
provide a benchmark for observability.

\begin{figure}
\includegraphics[width=7.5cm]{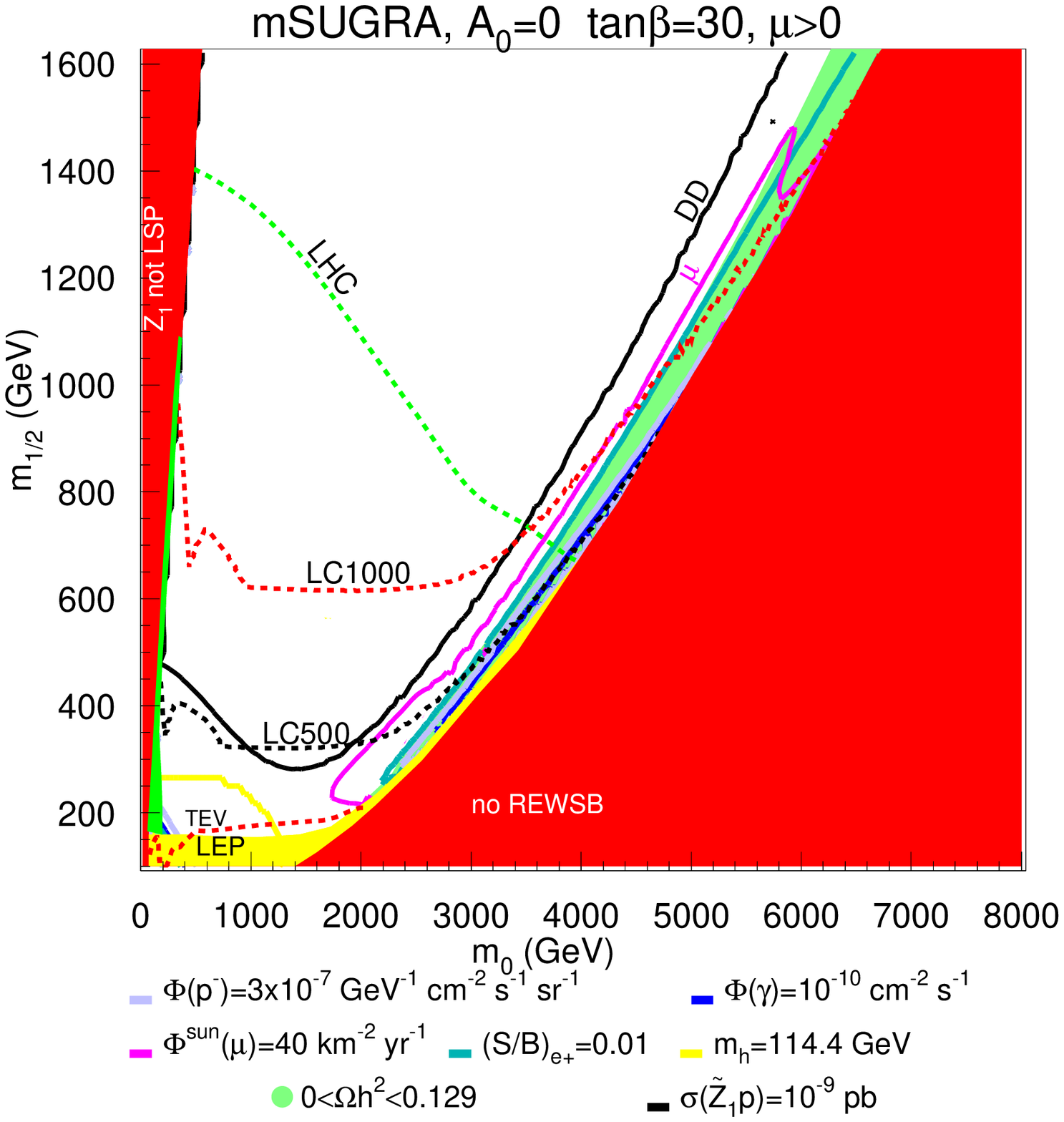}%
\includegraphics[width=7.5cm]{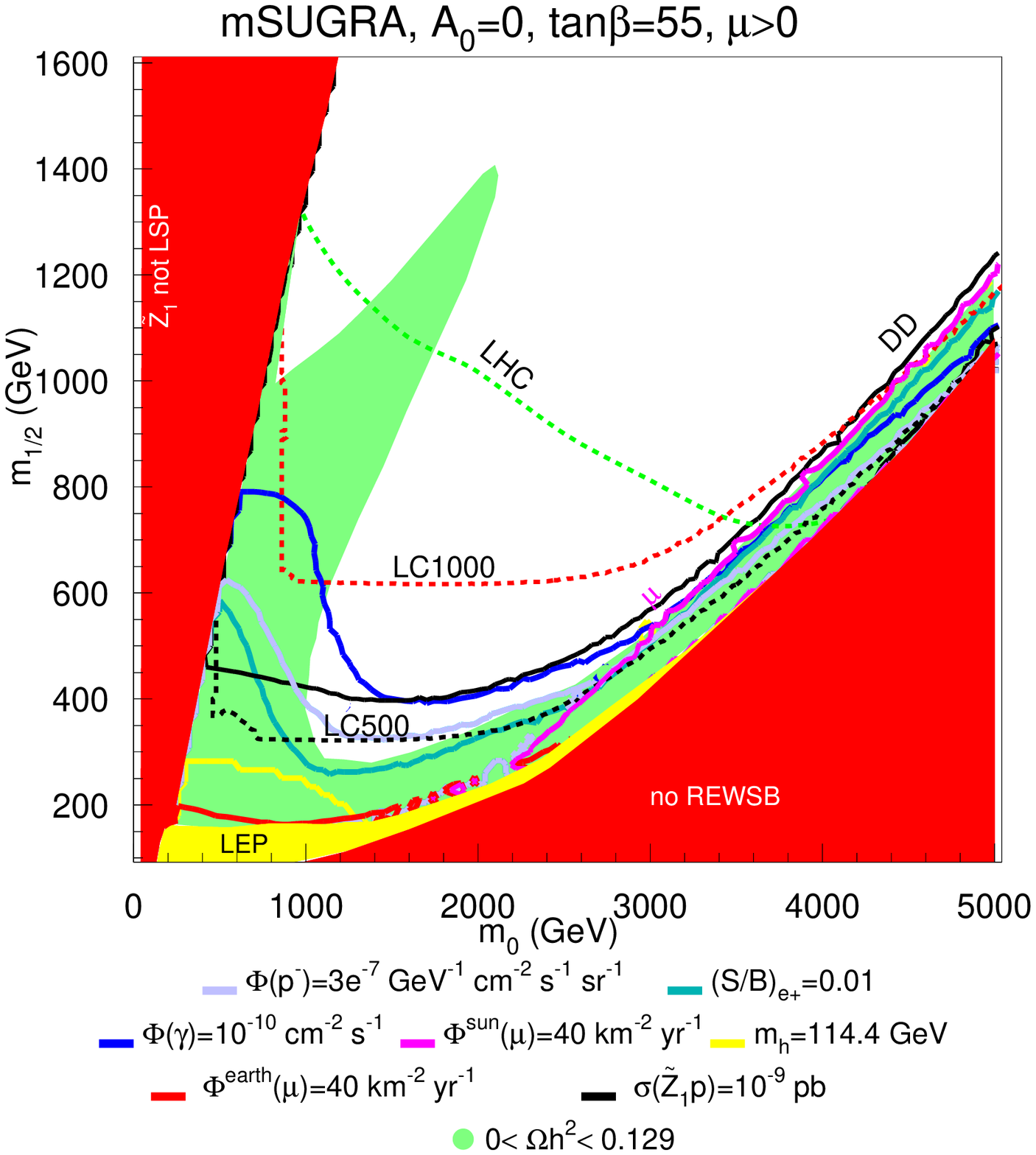}
\caption{\label{fig:dmsearch}A plot of the reach of direct, indirect and collider searches for 
neutralino dark matter in the $m_0\ vs.\ m_{1/2}$ plane, for 
$A_0=0$, $\tan\beta =30(55)$left(right) and $\mu >0$}
\end{figure}

Our results for direct/indirect DM search experimental 
reach for the mSUGRA model are shown in Fig. \ref{fig:dmsearch} in the 
$m_0\ vs.\ m_{1/2}$ plane for $\tan\beta =30$ and 55 (left and right respectively), 
$A_0=0$ and $\mu >0$.
If addition to collider reach presented in the previous figures, here w
e  show contours of 
\begin{itemize}
\item Stage 3 direct detection experiments
($\sigma_{SI}>10^{-9}$ pb; black contour),
\item reach of IceCube $\nu$ telescope with 
$\Phi^{sun}(\mu )=40\ \mu$s/km$^3$/yr and $E_\mu >25$ GeV
(magenta contour),
\item the $\Phi (\gamma)=10^{-10}\ \gamma$s/cm$^2$/s contour
with $E_\gamma >1$ GeV in a cone of 0.001 sr directed at the
galactic center (dark blue contour),
\item the $S/B>0.01$ contour for halo produced positrons
(blue-green contour) and
\item the anti-proton flux rate 
$\Phi (\bar{p})=3\times 10^{-7}\ \bar{p}$s/cm$^2$/s/sr 
(lavender contour).
\end{itemize}
As noted by Feng {\it et al.}~\cite{Feng:2000gh,Feng:2000zu}, 
{\it all} these indirect signals
are visible inside some portion of the HB/FP region, while {\it none} 
are visible in generic DM disallowed regions (under the assumed smooth
halo profiles).
The intriguing point is that almost the entire HB/FP region 
can be explored
by the cubic km scale IceCube $\nu$ telescope! It can also be explored 
(apparently at later times) by the Stage 3 direct DM detectors. 
Taking into account relative time scales of the various search
experiments, if SUSY lies within the upper HB/FP region, then
it could  be discovered first by IceCube (and possibly Antares),
with a signal being later confirmed by direct DM detection and possibly by the 
TeV scale linear $e^+e^-$ collider. There is also some chance to obtain
indirect $\gamma$, $e^+$ and $\bar{p}$ signals in this region.
Notice that if instead SUSY lies within the stau co-annihilation corridor, 
then it will be  discovered by the LHC, but all
indirect detection experiments will find null results in their DM searches.

\section{Beyond mSUGRA: normal mass hierarchy }

We see that  $\Delta a_\mu$ favors light second-generation
sleptons, while $BF (b\to s\gamma )$ prefers heavy third generation squarks.
Since this situation is hard to realize in the mSUGRA model, 
this could be an 
indication that one must move beyond the assumption of 
universality, wherein each generation has a common mass at
$Q=M_{GUT}$.
Therefore scenario in  which first and
second generation scalars remain degenerate, while
allowing for a significant splitting with third generation
scalars is well motivated.
In this case, heavy (multi-TeV) third generation
scalars are preferred by $BF(b\to s\gamma )$ constraints,
while rather light first and second  generation scalars are
preferred by $\Delta a_\mu$.
The scenario is called the
normal scalar mass hierarchy (NMH).
The parameter
set of the mSUGRA model is expanded to the following values:~\cite{Baer:2004xx}
\begin{equation}
\label{eq:nusugra-space}
m_0(1),\ m_0(3),\ m_H,\ m_{1/2},\ A_0,\ \tan\beta,\ sign (\mu ) .
\end{equation}
where $m_0(1)$ is the common scalar mass of all {\it first and second}
generation scalars at $Q=M_{GUT}$, while $m_0(3)$ is the common
mass of all {\it third} generation scalars at $M_{GUT}$. 
The above parameter set is well motivated
in $SO(10)$ SUSY GUT models, where the two MSSM Higgs doublets
typically occupy a {\bf 10} of $SO(10)$, and each generation
of scalars, along with a SM gauge singlet $N$ occupies the 
{\bf 16} dimensional spinorial representation of $SO(10)$.
The step of breaking generational universality must be 
taken with some caution, 
since in general it can lead to violations of constraints 
from FCNC processes.
Splitting the third generation from the first and second can 
potentially lead  to violations of FCNC processes.  One of the main
experimentally measured bounds on FCNC processes in this case comes from
$B^0_H-\bar{B^0_L}$ mass splitting. 
One can show that  $\Delta m_B$ 
is much less restrictive than the kaon case, for both low and high
squark masses. Moreover,
even if squark mass splittings are very large at the scale 
$Q=M_{GUT}$, the weak scale mass splittings are much smaller
and practically   whole parameter space of NMH SUGRA is allowed~\cite{Baer:2004xx}.

Scan in the SUGRA parameter space of Eq.~\ref{eq:nusugra-space}
and respective $\chi^2$ 
analysis show the following preferences~\cite{Baer:2004xx}:
$m_H\sim m_0(3)$, $m_0(1)\ll m_0(3)$,  $m_0(1)\sim 0-400$ GeV, 
$m_0(3)\sim 500-3000$ and large $m_0(1)-m_0(3)$ 
mass splitting and  that mainly
large $\tan\beta$ is preferred if the generational mass splitting
is small (which takes us back towards the mSUGRA case).
If may also be pointed out that SUSY IMH models
are greatly disfavored. The above scan motivates to reduce 
the number of parameters and choose $m_H=m_0(3)$ which brings us 
to the minimal extension of mSUGRA space by 
just one additional parameter $m_0\to [m0(1),m_0(3)]$.
\begin{figure}
\epsfig{width=7.1cm,file=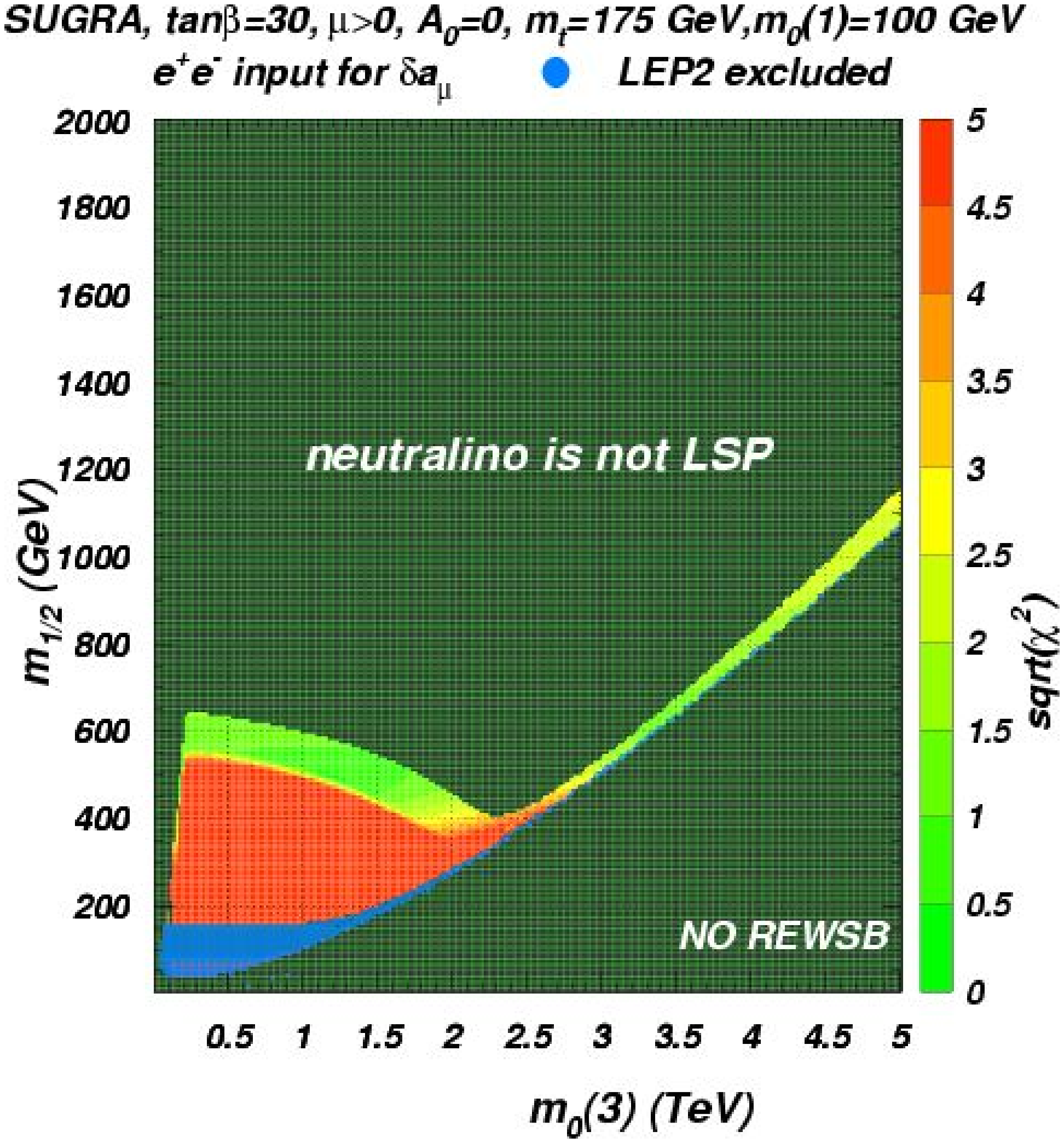}
\epsfig{width=7.cm,file=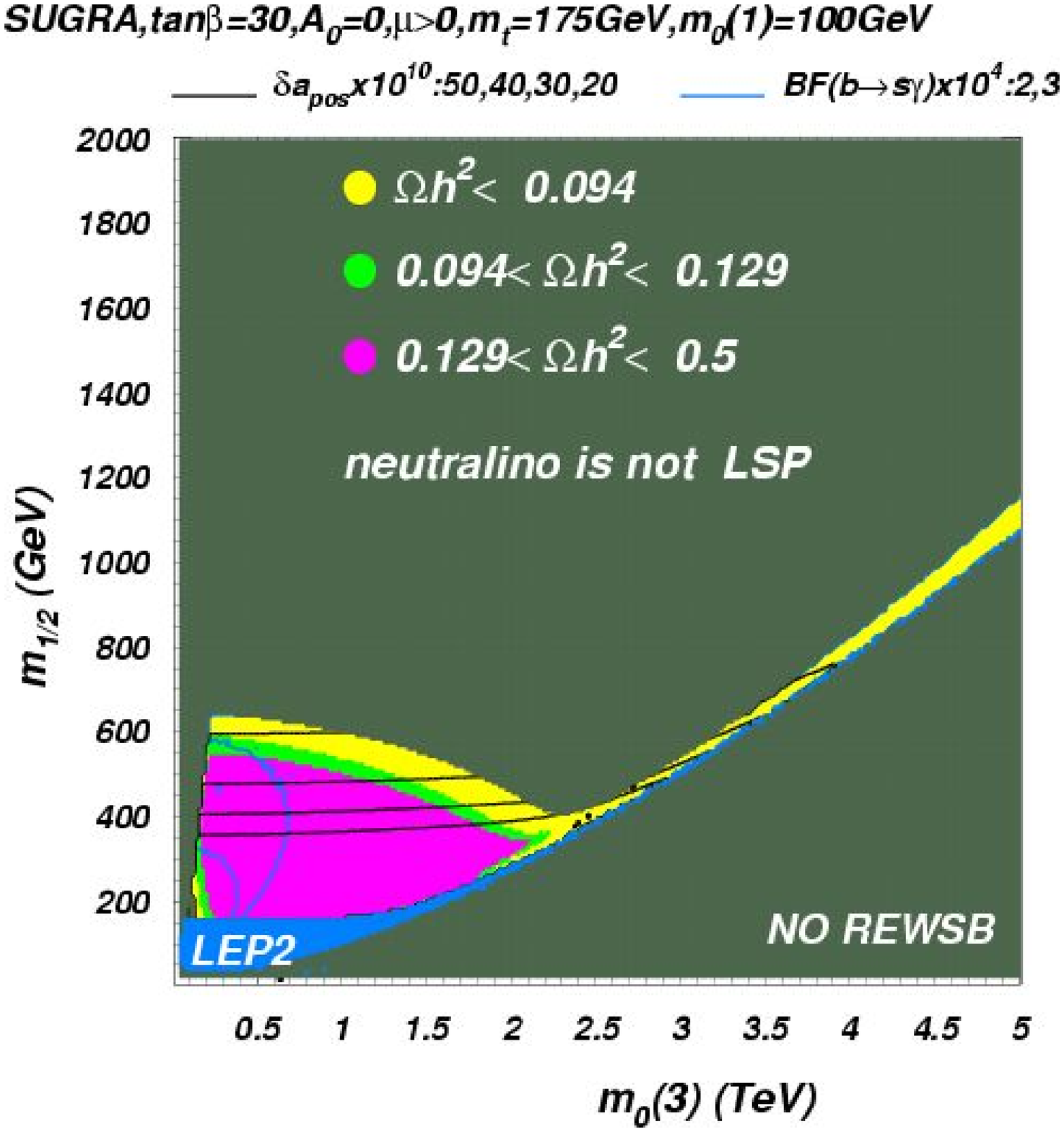}
\caption{\label{fig:nusug-chi2} 
Left frame:
$\sqrt{\chi^2}$ values
in the $m_0(3)\ vs.\ m_{1/2}$ plane 
for $m_0(1)=100$~GeV,  $\tan\beta =30$, $A_0=0$ and $\mu >0$
in NMH SUGRA scenario.The green regions have low $\chi^2$, while red regions have
high $\chi^2$. Yellow is intermediate.
Right frame:
constraints for
NMH SUGRA model in the $m_0(3)$ vs. 
$m_{1/2}$ plane for  $m_0(1)=100$GeV, $A_0=0$ and $\tan\beta=30$.
Allowed CDM relic density regions are the green shaded ones,
LEP2 excluded regions denoted by blue color.
Black contours denotes $\delta a_\mu$ 
values ($30,10,5,1\times 10^{-10}$ -- from bottom to top)
and blue contours denote $BF(b\to s\gamma)$ 
values ($2,3\times 10^{-4}$ -- from bottom to top).
The bottom raw of figures presents $\sqrt{\chi^2}$ 
for the respective parameters of the mSUGRA model}
\end{figure}
In Fig.~\ref{fig:nusug-chi2}(left) 
we present   the $\sqrt{\chi^2}$ values
in the $m_0(3)\ vs.\ m_{1/2}$ plane 
for $m_0(1)=100$~GeV,  $\tan\beta =30$, $A_0=0$ and $\mu >0$.
The corresponding contour plots of
$BF(b\to s\gamma)$, $a_\mu$ and $\Omega h^2$ are shown in 
right frame of the figure.
One can see that most of the area displayed is
excluded. In this case, slepton masses are quite light, in the vicinity
of a few hundred GeV. As $m_{1/2}$ increases, ultimately $m_{\tz_1}$
becomes greater than $m_{\tell}$, and one violates the cosmological constraint
on stable charged relics from the Big Bang.
On the other hand, of  the {\it surviving} parameter space, 
there are large regions with relatively low $\chi^2$. The plot with
$m_0(1)=100$ GeV has a rather broad band of low $\chi^2$. In this case,
neutralinos in the early universe can annihilate by a combination
of $t$-channel slepton exchange (as in the bulk region of mSUGRA), and
by neutralino-slepton co-annihilation. In addition, smuons and mu sneutrinos 
are relatively light, giving a large, positive contribution to
$\Delta a_\mu$, while top squarks inhabit the TeV and beyond range, 
effectively suppressing anomalous contributions to $BF(b\to s\gamma )$.
In addition to the region with low values of $m_0(3)$ and low $m_{1/2}$,
an important  portion the HB/FP region survives
since low $m_0(1)$ value
provides big enough contribution to $\Delta a_\mu$ 
while big $m_0(3)$ value keeps SUSY contribution to 
$BR(b\to s\gamma)$ suppressed.

\begin{floatingfigure}{8cm}
\hspace*{-1.2cm}\epsfig{file=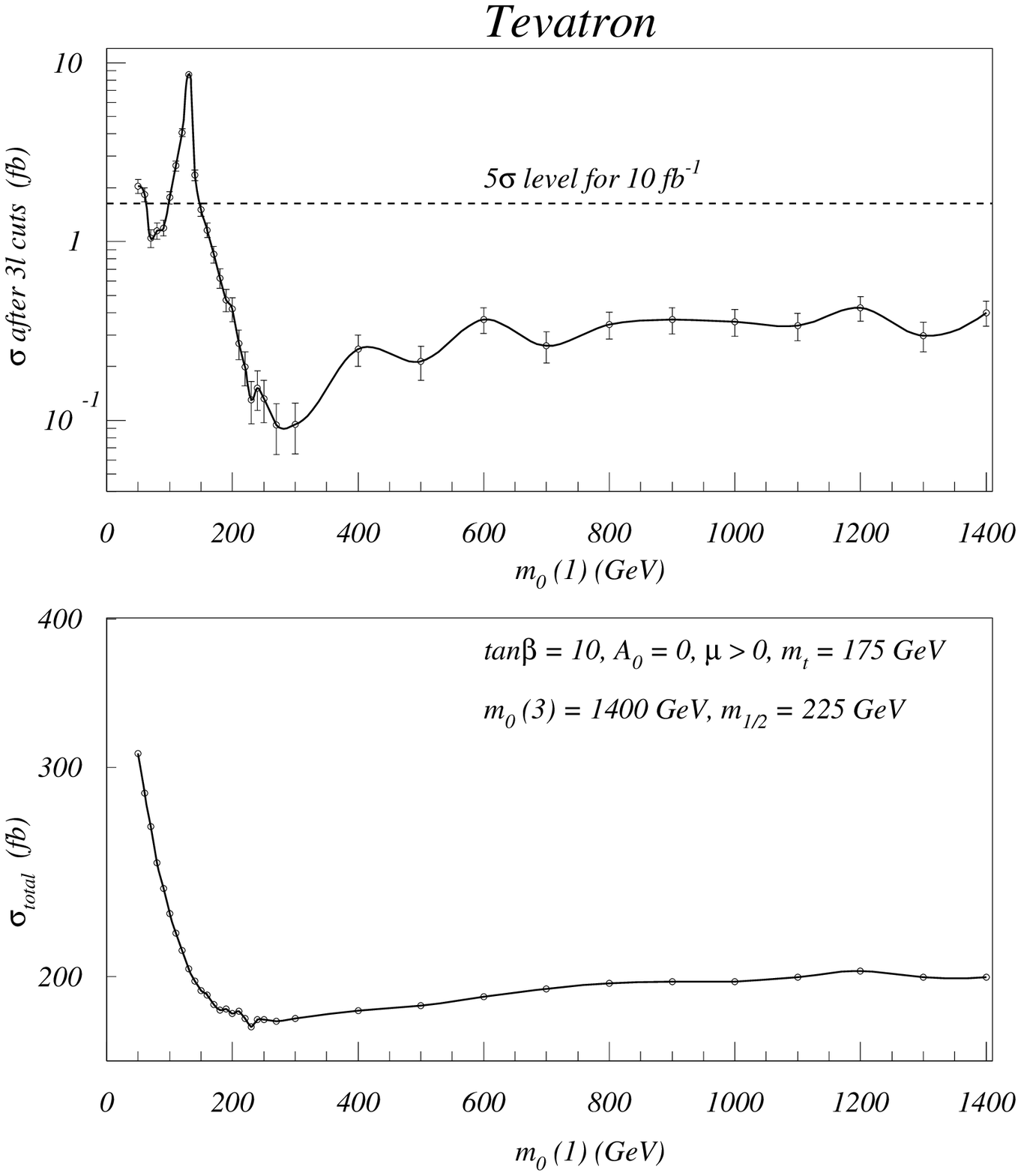,width=9cm}
\ffcaption{Rates for isolated trilepton events at the Fermilab Tevatron
$p\bar{p}$ collider, after cuts SC2 from 
Ref.~\cite{Matchev:1999yn,Baer:1999bq,Barger:1998hp}
}
\label{fig:tevatron}
\end{floatingfigure}

The scenario
of NMH SUGRA model works if $m_0(1)\simeq m_0(2)\ll m_0(3)$, and leads 
to spectra typically with squarks and third generation sleptons 
in the TeV range, while first and second generation sleptons 
have masses in the range of 100-300 GeV. The presence of rather
light first and second generation sleptons in the sparticle mass spectrum
in general leads to enhancements in leptonic cross sections from
superparticle production at collider experiments, compared to the case 
where selectrons and smuons are in the multi-TeV range --- 
sleptons may now be produced 
with non-negligible cross sections at colliders, and also in that
their presence enhances the leptonic branching fractions of charginos
and especially neutralinos. Therefore it is worth to discuss the implications of
light selectrons and smuons for the Fermilab Tevatron collider, the
CERN LHC and a linear $e^+e^-$ collides.

It is 
usually expected that $p\bar{p}\to \tw_1^+\tw_1^- X$ and
$\tw_1\tz_2 X$ will be the dominant production cross sections~\cite{Baer:1995nq}
at the Tevatron if sparticle are accessible.
If $\tw_1\to\ell\nu_\ell\tz_1$ and $\tz_2\to \ell\bar{\ell}\tz_1$, then
clean trilepton signals may occur at an observable rate. Signal
and background rates have recently been investigated 
in Ref.~\cite{Matchev:1999yn,Baer:1999bq,Barger:1998hp}, 
where Tevatron reach plots may also be found.
Fig.~11 presents isolated trilepton  Isajet 7.69 signal
after using cuts SC2 of Ref.~\cite{Baer:1999bq}, where the backgrounds are also
evaluated for for the
parameter space point $m_0(3)=1400$ GeV, $m_{1/2}=225$ GeV, $A_0=0$, 
$\tan\beta =10$ and $\mu >0$. 
The results are  plotted versus variation in the $m_0(1)$ parameter, 
the signal level needed
for a $5\sigma$ signal with 10 fb$^{-1}$ is also denoted. The error bars show 
the Monte Carlo statistical error. When 
$m_0(1)=m_0(3)$ (at $m_0(1)=1400$ GeV), the results correspond to the mSUGRA
model, and the isolated trilepton signal is well below discovery
threshold.
At $m_0(1)\lesssim 200$~GeV (which is favorable by $\delta a_\mu$ data)
the light sleptons begin to dominate
neutralino three body decays rates, and consequently 
the trilepton cross section rises steeply, to the level of observability. 
Eventually chargino and and neutralino two-body decays 
to sleptons turn on (in this case, first $\tz_2\to \tell_R\ell$), 
and trilepton rates become very high.

Similarly, at the the CERN LHC,
$m_0(1)$ drops below about 200~GeV (the value typically needed to 
explain the $(g-2)_\mu$ anomaly), multilepton rates rise steeply.
Thus, we would expect that SUSY -- as manifested in the NMH SUGRA model --
would be easily discovered, and what's more, the signal events would be
unusually rich in multilepton events. Such multilepton events can be 
especially useful for reconstructing sparticle masses in gluino and squark
cascade decay events~\cite{Hinchliffe:1996iu}.

A linear collider operating at $\sqrt{s}=0.5-1$ TeV may be the next frontier
particle physics accelerator beyond the CERN LHC. 
Depending on sparticle masses and the collider energy, 
charginos and neutralinos may or may not be accessible. However, 
in the NMH SUGRA model, light first and second generation sleptons are 
needed both to explain the $(g-2)_\mu$ anomaly, but also to enhance
neutralino annihilation in the early universe. This means slepton masses
are typically in the 100-300 GeV range, and likely within reach of 
a linear $e^+e^-$ collider~\cite{Baer:2003ru}.

\section{SUSY GUTs and Yukawa unified SUSY models}
The existence of the weak scale supersymmetry leads to gauge coupling unification
at the scale of $M_{GUT}\simeq 2\time 10^{16}$~GeV which in its turn
is compelling hing for SUSY grand unified theories or SUSY GUTs.
Such a 
SUSY GUT theory may be the ``low energy effective theory'' that can
be obtained from some more fundamental superstring theory. 
Models based on the gauge group $SU(5)$ are compelling in that they 
explain the apparently ad-hoc
hypercharge assignments of the SM 
fermions~\cite{Georgi:1974yf,Georgi:1974sy,Buras:1977yy}. 
However, many $SU(5)$ SUSY GUT models
as formulated in four dimensions are already excluded by proton 
decay constraints~\cite{Murayama:2001ur}.
$SU(5)$ SUSY GUT models can also be formulated in five or more dimensions,
where compactification of the extra dimensions leads to a break down 
in gauge symmetry~\cite{Hebecker:2001wq,Hall:2001pg,Kobakhidze:2001yk,%
Altarelli:2001qj,Kawamoto:2001wm}. These models can dispense with the unwieldy 
large Higgs representations required by four-dimensional models, and can also
be constructed to suppress or eliminate proton decay entirely.

Further step to the complete   gauge and matter field  family unification (as
well as Yukawa coupling unification as we discuss below) can be achieved in
SUSY GUTs models based on SO(10) group which therefore looks 
very 
intriguing~\cite{Mohapatra:1999vv,Gell-Mann:1976pg,Fritzsch:1974nn,%
Raby:2004br,Altarelli:2004za}.
In addition to unifying gauge couplings,   
\begin{itemize}  
\item They unify all matter of a single generation into the  
16 dimensional spinorial multiplet of $SO(10)$.  
\item The {\bf 16} of $SO(10)$ contains in addition to all SM  
matter fields of a single generation a gauge singlet   
right handed neutrino state which naturally leads to a mass  
for neutrinos. The well-known see-saw 
mechanism~\cite{Gell-Mann:1980vs,Yanagida:1979as,Mohapatra:1979ia}   
implies that if $m_{\nu_\tau}  
\sim 0.03$ eV, as suggested by atmospheric neutrino 
data~\cite{Fukuda:1998ah,Fukuda:2000np},   
then the mass scale associated with $\nu_R$ is very close  
to the $GUT$ scale: {\it i.e.} $M_N\sim 10^{15}$ GeV.  
\item $SO(10)$ explains the apparently fortuitous   
cancellation of triangle anomalies within the SM.  
\item The structure of the neutrino sector of $SO(10)$ models lends  
itself to a successful theory of baryogenesis via   
intermediate scale 
leptogenesis~(For a review, see~\cite{Buchmuller:2002xm}.  
\item
In the simplest $SO(10)$ SUSY GUT models, the two Higgs doublets 
of the MSSM occupy the same
10 dimensional Higgs multiplet $\phi({\bf 10})$. The superpotential 
then contains the term
$\hat{f}\ni f \hat{\psi}({\bf 16})\hat{\psi}({\bf 16})\hat{\phi}({\bf 10}) 
+\cdots$
where $f$ is the single Yukawa coupling for the third generation.
Thus, the simplest $SO(10)$ SUSY GUT models predict 
$t-b-\tau$ Yukawa coupling unification 
in addition to gauge coupling unification. 
\end{itemize}

It is possible to calculate the $t$, $b$ and $\tau$ Yukawa couplings at
$Q=m_{weak}$, and extrapolate them to $M_{GUT}$ in much the same way
one checks the theory for gauge coupling unification. Yukawa coupling
unification turns out to depend on the entire spectrum of 
sparticle masses and mixings
since these enter into the weak scale supersymmetric threshold corrections.
Thus, the requirement of $t-b-\tau$ Yukawa 
coupling unification can be a powerful constraint on the 
soft SUSY breaking terms of the low energy effective 
theory~\cite{Ananthanarayan:1991xp,Anderson:1993fe,Carena:1994bv,%
Rattazzi:1995gk,Ananthanarayan:1994qt,Blazek:1999ue}.

In Ref.~\cite{Baer:1999mc},
it was found that models with Yukawa coupling unification with $R<1.05$
(good to 5\%) 
could be found if additional $D$-term splittings of scalar masses
were included. 
Here, $R\equiv max(f_t,\ f_b,\ f_\tau )/min (f_t,\ f_b,\ f_\tau )$, where
all Yukawa couplings are evaluated at the GUT scale.
The $D$-term splittings occur naturally when the 
$SO(10)$ gauge symmetry breaks to $SU(5)$, and they are given by
(see~\cite{Auto:2004km} and references in it).
\begin{eqnarray}  
m_Q^2=m_E^2=m_U^2=m_{16}^2+M_D^2 , \nonumber \\  
m_D^2=m_L^2=m_{16}^2-3M_D^2 , \nonumber \\  
m_N^2 = m_{16}^2+5M_D^2,\nonumber \\  
m_{H_{u,d}}^2=m_{10}^2\mp 2M_D^2 ,  
\label{eq:so10}
\end{eqnarray}  
where $M_D^2$ parameterizes the magnitude of the $D$-terms.  
Owing to our ignorance of the gauge symmetry breaking mechanism,  
$M_D^2$ can be taken as a free parameter,   
with either positive or negative values.  
$|M_D|$ is expected to be of order the weak scale.  
Thus, the $D$-term ($DT$) model is characterized by the following free   
parameters,  
\begin{eqnarray}  
m_{16},\ m_{10},\ M_D^2,\ m_{1/2},\ A_0,\ \tan\beta ,\ {\rm sign}(\mu ).  
\label{eq:pspace}
\end{eqnarray}

Using the $DT$ model, Yukawa unification good to 5\% was found
when soft term parameters $m_{16}$ and $m_{10}$ were scanned up to
1.5 TeV, but only for $\mu <0$ values~\cite{Baer:1999mc,Baer:2000jj}. 
The essential quality of the $D$-term 
mass splitting is that it gave the value of $m_{H_u}$ a head start 
over $m_{H_d}$ in running towards negative values, as is required for REWSB.
Good relic density was also found in the stau co-annihilation and 
hyperbolic branch/focus point (HB/FP) region, 
but at some cost to the degree of
Yukawa coupling unification.

 In Ref.~\cite{Baer:2001yy}, it was found that Yukawa coupling 
unification good to only 30\% could be achieved in $DT$ models with
$\mu >0$ when $m_{16}$ values were scanned up to 2 TeV. The models with the
best Yukawa coupling unification were found to have soft term 
relations
\be
A_0^2\simeq 2 m_{10}^2\simeq 4 m_{16}^2 ,
\ee
which had also been found by Bagger {\it et al.} in the context of
radiatively driven inverted scalar mass hierarchy (IMH) 
models~\cite{Feng:1998iq,Bagger:1999ty,Bagger:1999sy,%
Baer:1999md,Baer:2001vw,Baer:2000gf}.

In Ref.~\cite{Baer:2001yy}, a model with just GUT scale Higgs mass splittings was
also examined ($HS$), while all other scalars remained universal.

The parameter space of the $HS$ model is that of Eq.~\ref{eq:pspace}, 
but where the $D$-term splitting is only applied to the last 
of the relations in Eq.~\ref{eq:so10}.
Yukawa coupling unification in the $HS$ model was found to be comparable 
to the $DT$ model case when $m_{16}$ values up to 2 TeV were scanned.
In Ref.~\cite{Auto:2003ys} (Auto {\it et al.}), 
soft term values of $m_{16}$ up to
20 TeV were explored. In this case, Yukawa unified solutions to better 
than 5\% were found for $\mu >0$ for the $HS$ model when very large 
values of $m_{16}>5-10$~TeV were scanned. The large scalar masses
that gave rise to Yukawa unification also acted to suppress neutralino
annihilation in the early universe, so that rather large values of 
the relic density were found. Models with $\Omega_{\tz_1}h^2<0.2$ could be
found, but only at the expense of accepting Yukawa coupling 
unification to 20\%, rather than 5\%. The models found with low relic density
generally had either a low $\mu$ value, or were in the light Higgs
annihilation corridor, with $2m_{\tz_1}\sim m_h$.

\begin{floatingfigure}{7cm}
\hspace*{-1cm}%
\includegraphics[width=8cm]{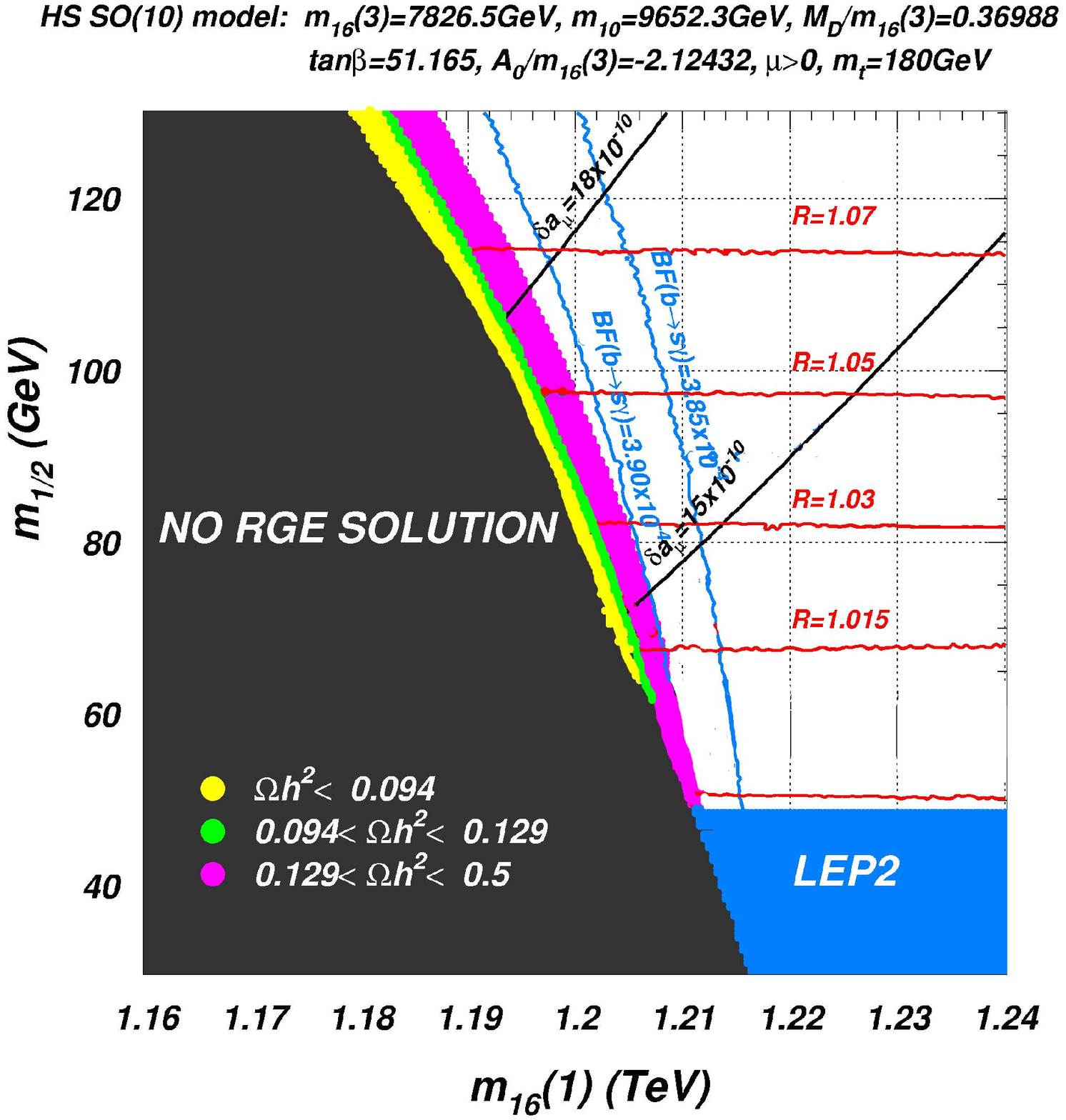}
\ffcaption{
\label{fig:so10}
Allowed parameter space of Yukawa unified  supersymmetric 
HS model with generational non-universality from~\cite{Auto:2004km}.
We show the
$m_{16}(1)\ vs.\ m_{1/2}$ plane for $m_{16}(3)=7830$ GeV, 
$m_{10}=9650$ GeV, $M_D/m_{16}(3)=0.37$, $A_0/m_{16}(3)=-2.1$, 
$\mu >0$, $\tan\beta =51$
and $m_t=180$ GeV.
The black shaded region gives tachyonic particles, while the blue region is 
excluded by LEP2 chargino search experiments. The yellow and green
regions are allowed by the WMAP determination of $\Omega_{\tz_1}h^2$. We also
show contours of $R$, the measure of Yukawa unification at $M_{GUT}$.
}
\label{fig-susy-papers}
\end{floatingfigure}

Similar work has been carried on by Blazek, Dermisek and Raby (BDR).
In Ref.~\cite{Blazek:2001sb,Blazek:2002ta}, 
the BDR group used a top-down approach to the RG
sparticle mass
solution to find Yukawa unified solutions for $\mu >0$, where they also noted 
that in this case the $HS$ model worked better than the $DT$ model.
In their approach, the third generation fermion masses and other
electroweak observables were an output of the program, so that starting
with models with perfect Yukawa coupling unification, 
they would look for solutions with a low $\chi^2$ value 
constructed from the low energy observables. The BDR Yukawa unified solutions
were also characterized by soft term IMH model boundary conditions.
The solutions differed from those of Ref.~\cite{Auto:2003ys} in that 
they always gave a very low value of Higgs mass $m_A$ and also small $\mu$
parameter, indicative of a mixed higgsino-bino LSP. In Ref.~\cite{Dermisek:2003vn},
the neutralino relic density was examined for the BDR solutions.
Their low $\mu$ and $m_A$ values generally led to very low values of
$\Omega_{\tz_1}h^2$ unless $m_{1/2}$ was small enough compared to 
$\mu $ that the LSP was in the mixed higgsino-bino region.

Finally, one more step was done in~\cite{Auto:2004km}
to solve the problem of tension between Yukawa unification 
and low relic density in SO(10) SUSY GUTs models.
In this paper, two methods were explored to reconcile
Yukawa unified sparticle mass solutions with the neutralino relic density.
The first case is to allow splitting of the third generation of scalars
from the first two generations. By decreasing the first and second generation
scalar masses, sparticle mass solutions with very light
$\tu_R$ and $\tc_R$ squark masses, in the 90-120 GeV range
were obtained.
The other examined scenario  was the case of non-degenerate gaugino masses
at the $GUT$ scale. Beginning with any of the Yukawa unified
solutions found in an earlier study, it was found  that by dialing $M_1$ to
large enough values, a (partially) wino-like LSP can be generated, with a 
relic density in accord with WMAP allowed values.

In Fig.~12, we show the example 
of such solution for the first scenario
in the $m_{16}(1)\ vs.\ m_{1/2}$ 
plane for fixed $m_{16}(3)=7826.5$ GeV.
The blue region at low $m_{1/2}$ is excluded
by the LEP2 bound on chargino mass $m_{\tw_1}>103.5$ GeV. The black region 
on the left is excluded because the $m_{U_1}^2$ squared mass 
is driven tachyonic,
resulting in a color symmetry violating ground state. The green region
denotes sparticle mass spectrum solution with relic density
$0.094<\Omega_{\tz_1}h^2< 0.129$, within the WMAP favored regime, while the
yellow region denotes even lower relic density values, wherein
the CDM in the universe might be a mixture of neutralinos plus some other
species. The red region has $\Omega_{\tz_1}h^2<0.5$.
The remaining unshaded regions give sparticle spectra solutions with
$\Omega_{\tz_1}h^2>0.5$, and would be excluded by WMAP. We also show several
contours of $BF(b\to s\gamma )$, $\Delta a_\mu^{SUSY}$ and $R$, the
$max/min$ ratio of GUT scale Yukawa couplings. We see that all constraints
on the sparticle mass spectrum are within allowable limits for the
shaded region to the right of the excluded region.

The origin of 
the light squarks comes from the $S$ term in the one-loop RGEs, which 
is non-zero and large in the case of multi-TeV valued split Higgs masses.
A search for just two light squarks at the Fermilab Tevatron collider
using the new Run 2 data should be able to either verify or disprove this
scenario. In addition, the light squarks give rise to large rates for
direct detection of dark matter, and should give observable rates
for searches at CDMS II.

\vspace*{-0.5cm}

\section{Conclusions}

Author would like to apologize in advance for leaving several other interesting SUSY related topics
outside the scope of this review. Among them are CP and flavor violating SUSY physics,
SUSY R-parity violation, beyond the MSSM (nMSSM) scenarios and  various SUSY-related
aspects of baryogenesis. There are also  open fundamental problems,
which still need to be solved. One of the most important problems  is 
Cosmological constant problem and the  problem of origin of the $\mu$ term.

The most of the attention in this review  was paid to SUGRA models which 
are  very compelling indeed.
In constraining SUGRA parameter space CDM constraints play a crucial role
leaving only a few restricted regions to be tested by other experiments.
LEP2+($b\to s\gamma$)+($g-2$) constraints leave
{\it focus point}, {\it funnel} and {\it stau-co-annihilation}
region survived. LHC  collider can uniquely cover
the funnel region and (almost) all stau-co-annihilation region,
but leaves most of the focus point region uncovered.
On the other hand,  NLC could greatly extend LHC reach in focus point region.
There is also a great complementary role
of direct and indirect DM search experiments which has the potential
to cover completely focus point region leaving 'no escape'
for mSUGRA with upcoming experiments combined!

There could be an indication coming from  from $\delta a_\mu$ data that 
normal mass hierarchy in SUGRA scenario could be preferred by nature
and that, may be, mSUGRA scenario is too simple to be true.

Talking about  GUTs models, SO(10) SUSY GUTs
looks very attractive from both, theoretical and phenomenological points of view,
and predicts very specific particle spectra testable experimentally.

Finally, I would like to express my belief that
the our exciting  era of upcoming experiments will be indeed successful
in hunting for Supersymmetry which could be just around corner!

%
%

\vspace*{-0.5cm}

\begin{theacknowledgments}
Author would like to thank his FSU colleagues,
Howie Baer, Daniel Auto, Azar Mustafayev,  Jorge O'Farrill, Tadas Krupovnikas
and Csaba Balazs for very fruitful  collaboration.
I am also grateful to Dmitri Kazakov, Alexei Gladyshev, Daisuke Nomura, Sekhar Shivukula,
Elizabeth Simmons, Anatoly Solomin, Kazuhiro Tobe,   Wu-Ki Tung and C.-P. Yuan
for numerous stimulating  discussions during the preparation of this review.
DOE support is also acknowledged.
\end{theacknowledgments}


\bibliographystyle{aipproc}   

\bibliography{hcp04_belyaev}

\end{document}